\documentclass[aps,showpacs,superscriptaddress]{revtex4}
\usepackage[dvips]{graphicx}
\usepackage{epsfig}
\usepackage{amsmath,amssymb}
\numberwithin{equation}{section}
\newcommand{\dbarp}{\frac{d^3p}{(2\pi)^3}}
\newcommand{\dbarq}{\frac{d^3q}{(2\pi)^3}}
\newcommand{\ddbarp}{\frac{d^dp}{(2\pi)^d}}
\newcommand{\ddbarq}{\frac{d^dq}{(2\pi)^d}}
\newcommand{\ddbark}{\frac{d^dk}{(2\pi)^d}}

\begin{document}
\title{Dynamics near the critical point: the hot renormalization group\\in quantum field theory}
\author{D. Boyanovsky}
\email{boyan@pitt.edu} \affiliation{Department of Physics and
Astronomy, University of Pittsburgh, Pittsburgh, Pennsylvania
15260, USA}
\author{H. J. de Vega}
\email{devega@lpthe.jussieu.fr}
 \affiliation{LPTHE, Universit\'e
Pierre et Marie Curie (Paris VI) et Denis Diderot (Paris VII),
Tour 16, 1er. \'etage, 4, Place Jussieu, 75252 Paris, Cedex 05,
France}

\date{\today}

\begin{abstract}
The perturbative approach to the description of long wavelength
excitations at high temperature breaks down near the critical
point of a second order phase transition. We study the
\emph{dynamics} of these excitations  in a relativistic scalar
field theory at and near the critical point via a renormalization
group approach at high temperature and  an  $\epsilon$ expansion
in $d=5-\epsilon$ space-time dimensions. The long wavelength
physics is determined by a non-trivial fixed point of the
renormalization group. At the critical point we find that the
dispersion relation and width  of quasiparticles of momentum $p$
is $\omega_p \sim p^{z}$ and $\Gamma_p \sim (z-1) \, \omega_p$
respectively, the group velocity of quasiparticles $v_g \sim
p^{z-1}$ vanishes in the long wavelength limit at the critical
point. Away from the critical point for $T\gtrsim T_c$ we find
$\omega_p \sim \xi^{-z}\left[1+(p\,\xi)^{2z}\right]^{\frac{1}{2}}$
and $\Gamma_p \sim (z-1)~\omega_p~
 \frac{(p\,\xi)^{2z}}{1+(p\,\xi)^{2z}}$ with $\xi$ the
 finite temperature correlation length $ \xi \propto
 |T-T_c|^{-\nu}$. The new \emph{dynamical} exponent $z$
 results from anisotropic renormalization in the spatial and
time directions. For a theory with $O(N)$ symmetry we find
 $z=1+ \epsilon ~ \frac{N+2}{(N+8)^2}+\mathcal{O}(\epsilon^2)$.This dynamical critical exponent describes a new
universality class for dynamical critical phenomena in quantum
field theory.
  Critical slowing down, i.e, a vanishing width in
the  long-wavelength limit, and the validity of the
 quasiparticle picture emerge naturally from this analysis.
\end{abstract}


\maketitle
\section{\label{sec:intro} Introduction}
The experimental possibility of studying the phase transitions of
QCD via ultrarelativistic heavy ion collisions with the current
effort at RHIC and the forthcoming program at LHC motivates a
theoretical effort  to understand the dynamical aspects of phase
transitions at high temperature. QCD is conjectured to feature two
phase transitions, the confinement-deconfinement (or
hadronization) and the chiral phase transitions. Detailed lattice
studies\cite{karsch} seem to predict that both transitions occur
at about the same temperature $T_c \sim 170 \mathrm{Mev}$.

While lattice gauge theories furnish a non-perturbative tool to
study the thermodynamic equilibrium aspects of the transition the
\emph{dynamical} aspects cannot be accessed with this approach.

In a condensed matter experiment  the temperature is typically a
control parameter and it can be varied sufficiently slow so as to
ensure that a phase transition occurs in local thermodynamic
equilibrium. In an ultrarelativistic heavy ion collision the
current theoretical understanding suggests that a thermalized
quark-gluon plasma may be formed at a time scale of order
$1\,\mathrm{fm}/c$ with  a temperature larger than the critical.
This quark-gluon plasma then expands hydrodynamically and cools
almost adiabatically, the temperature falling off as a power of
time $T(t) \sim T_i (t_i/t)^{1/3}$ until the transition
temperature is reached at a time scale $\sim 10-50\,
\mathrm{fm}/c$ depending on the initial temperature\cite{qgp}.

Whether the phase transition occurs in local thermodynamic
equilibrium or not depends on the ratio of the cooling time scale
$t_{cool} \sim T(t)/\dot{T}(t)$ to the relaxation or
thermalization time scale of a fluctuation of a given wavelength
$p^{-1}$, $t_{rel}(p)$. If $t_{cool} \gg t_{rel}$ then the
fluctuation relaxes in time scales much shorter than that of the
temperature variation and reaches local thermodynamic equilibrium.
If, on the other hand, $t_{cool} \ll t_{rel}$ the fluctuation does
not have time to relax to local thermodynamic equilibrium and
freezes out. For these fluctuations the phase transition occurs
\emph{very fast} and out of equilibrium. Thus an important
\emph{dynamical} aspect  is  to understand the relaxation time
scales for fluctuations.

A large body of theoretical, experimental and numerical work in
condensed matter physics reveal that while typically short
wavelength ($p \gg T$) fluctuations reach local thermal
equilibrium, near a critical point long wavelength fluctuations
relax very slowly, and undergo \emph{critical slowing
down}\cite{tdgl,ma}. A phenomenological description of the
dynamics near a phase transition typically hinges on the
time-dependent Landau-Ginzburg equation which is generalized to
include conservation laws\cite{tdgl,ma}. In the simplest case of a
non-conserved order parameter, such as in a scalar field theory
with discrete (Ising-like) symmetry,  the time dependent Landau
Ginzburg equation is purely dissipative.

 While phenomenological,
this approach has proved very succesful in a variety of
experimental situations and is likely to provide a suitable
description of the dynamics for macroscopic, coarse-grained
systems such as binary mixtures, etc.\cite{tdgl}. The
phenomenological approach based on the time dependent
Landau-Ginzburg equations which are first order in time
derivatives seem to provide a suitable description of
coarse-grained macroscopic dynamics in \emph{non-relativistic
systems}. However it is clear that such approach is not justified
in a relativistic quantum field theory, since the underlying
equations of motion are second order and time reversal invariant.

In particular in the case of purely dissipative time dependent
Landau Ginsburg (phenomenological) description\cite{tdgl,ma},
frequency and momenta enter with different powers in the
propagators, and at the mean field (or tree) level, this results
in a dynamical scaling exponent $z=2$. This situation must be
contrasted to that of a relativistic  quantum field theory where
at tree level (mean field) frequencies and momenta enter with the
same power in the propagator leading to a dynamical scaling
exponent $z=1$. Furthermore, critical slowing down is
automatically built in the phenomenological description, even at
\emph{tree level} as a consequence of dissipative equations of
motion\cite{spino}. Clearly this is not the case in a relativistic
quantum field theory. For a detailed discussion of the differences
of non-equilibrium dynamical aspects between the time dependent
Landau Ginzburg approach and quantum field theory see
ref.~\cite{spino}.

There are important non-equilibrium consequences of slow dynamics
near critical points. If the cooling time scale is much shorter
than the relaxation time scale of long-wavelength fluctuations,
these freeze out and undergo spinodal instabilities when the
temperature falls below the critical\cite{tdgl,spino} during
continuous (no metastability) phase transitions. These
instabilities result in the formation of correlated domains that
grow in time\cite{tdgl,spino} with a law that in general depends
on the cooling rate\cite{bowick}.

In ultrarelativistic heavy ion collisions during the expansion of
the quark gluon plasma the  critical point for the chiral phase
transition may be reached. If long wavelength fluctuations freeze
out shortly before the transition, the ensuing instabilities may
lead to distinct event by event observables\cite{rajagopal} in the
pion distribution as well as in the photon spectrum at low
energies\cite{photon}.

Thus an important aspect of the chiral phase transition is to
establish the relaxation time scales of long-wavelength
fluctuations, and whether critical slowing down and freeze out of
long-wavelength fluctuations can ensue.

 In the strict chiral limit with
massless up and down quarks, QCD has an $SU(2)_R \otimes SU(2)_L$
symmetry which is spontaneously broken to $SU(2)_{R+L}$ at the
chiral phase transition, the three pions being the Goldstone
bosons associated with the broken symmetry. It has been argued
that the low energy theory that describes the chiral phase
transition is in the same universality class as the Heisenberg
ferromagnet, i.e, the $O(4)$ linear Sigma model\cite{wilraja}.
This argument has been used\cite{wilraja} to provide an assessment
of the dynamical aspects of low energy QCD based on the
phenomenological time-dependent Landau Ginzburg approach to
dynamical critical phenomena in condensed matter\cite{tdgl}. While
the universality arguments are appealing, a more microscopic
understanding of dynamical critical phenomena in \emph{quantum
field theory} is needed and has begun to emerge only
recently\cite{pietroni,relaxationI}.

\vspace{3mm}

In reference\cite{pietroni} a Wilsonian renormalization group
extended to finite temperature was implemented in a scalar quartic
field theory. In this approach only one loop diagrams enter in the
computation of the beta functions, and the imaginary part of the
self energy, which arises first at two loops order for $T >T_c$ is
 accounted for by an imaginary part in the effective
quartic coupling\cite{pietroni}. There it is found that the
relaxation rate of zero momentum fluctuations $\gamma$ reveals
critical slowing down in the form $\gamma \sim
|T-T_c|^{\nu}\ln|T-T_c|$ with $ \nu \sim 0.53$ being the critical
exponent for the correlation length\cite{pietroni}.

In reference\cite{relaxationI}  the width of quasiparticles near
the critical point has been studied via the large N approximation.
This study revealed that at high temperature the effective
coupling is driven to a (Wilson-Fischer) fixed point, a result
that is in agreement with the numerical evidence presented in
reference\cite{pietroni}. While the results in leading order in
the large N limit found in\cite{relaxationI} hinted at critical
slowing down, albeit in a manner different from the numerical
evidence of reference\cite{pietroni}, they also hinted at the
breakdown of the quasiparticle picture. A conclusion
in\cite{relaxationI} is that while the large N limit provides a
partial resummation of the perturbative expansion, further
resummation is  needed to fully address the relaxation of
quasiparticles.

The large N limit in \emph{static} critical phenomena presents a
similar situation: while it sums the series of bubbles replacing
the bare vertex by the effective coupling that is driven to the
fixed point in the infrared, the self-energy still features
infrared logarithms that require further resummation\cite{ma}.
Such a resummation is provided by the renormalization
group\cite{ma}.

While our motivation for studying dynamical critical phenomena
near critical points is driven by the experimental program in
ultrarelativistic heavy ion collisions to study the QCD phase
transitions, the underlying questions are more overarching and of
a truly interdisciplinary nature. In particular we mention an
impressive body of work on aspects of \emph{quantum phase
transitions} in condensed matter systems\cite{sachdev} that
addresses very similar questions. The  work in
reference\cite{sachdev} focuses on understanding the static,
dynamical and transport properties of low dimensional systems in
the quantum regime, in which the frequency and momentum of
excitations is $\omega\,;\,p \gg T$.

Our study in this article is complementary to that program in that
we focus on the dynamical aspects of long-wavelength
quasiparticles with $\omega\,;\,p \ll T$.  As discussed
in\cite{sachdev,classical} and in detail below this is closer to
the \emph{classical} regime.

\vspace{3mm}

{\bf The goals:} in this article we study the \emph{dynamical}
aspects of quasiparticles near the critical point in a scalar
quartic field theory by implementing a renormalization group
program at high temperature. While the renormalization group has
been generalized to finite temperature in various
formulations\cite{senben,irla}, mainly to study  critical
phenomena associated with finite temperature phase transitions in
field theory, only \emph{static} aspects were studied with these
approaches.

Instead we focus on \emph{dynamical} aspects, in particular the
dispersion relations, and relaxation rates of long-wavelength
excitations  at and near the critical point. Already at the
technical level one can see the differences:  to understand
dynamical aspects, in particular relaxation, a consistent
treatment of absorptive parts of self-energies is required. This
aspect is notoriously difficult to implement in a Wilsonian
approach in Euclidean field theory\cite{senben}.
Reference\cite{pietroni} proposes a method to circumvent this
problem but a complete treatment that manifestly includes
absorptive parts of self-energy contributions is still lacking in
this approach. Other approaches using the Euclidean version of the
renormalization group adapted to finite temperature field theory
were restricted to static quantities\cite{irla} and in fact, as it
will be seen in detail below, miss important phenomena that will
be at the heart of the results presented here.

\vspace{3mm}

{\bf Brief summary of results:} Long-wavelength phenomena at high
temperature $T$ implies a dimensional reduction from the
decoupling of Matsubara modes with non-zero
frequency\cite{dimred1,dimred2}. The coupling in the dimensionally
reduced theory is $\lambda T$, where $\lambda$ is the quartic
coupling. By dimensional reasons, the perturbative expansion in
four space-time dimensions is in terms of the dimensionless ratio
$\lambda T/\mu$ with $\mu$ the typical momentum scale, which is
strongly relevant in the infrared. As a result, a perturbative
approach to studying long-wavelength phenomena breaks down. This
is manifest in the breakdown of the quasiparticle picture in naive
perturbation theory (see\cite{relaxationI} and below).

In $5-\epsilon$ space-time dimensions, the effective coupling in
the high temperature, long wavelength limit is $g(\mu)=\lambda T
\mu^{-\epsilon}$.  We implement an $\epsilon$ expansion around
\emph{five} space-time dimensions and a renormalization group
resummation program at high temperature with $T \gg s,\mu$ near
the critical point, with $s,\mu$ the typical frequency and
momentum scales. We analyze the high temperature behavior of the
relevant graphs and find that it is dominated for $ \epsilon > 0 $
by the zero Matsubara mode while the sum of the nonzero modes
gives subdominant contributions. The effective renormalized
coupling is driven to an infrared stable fixed point $ g^* = {\cal
O}(\epsilon) $, which for small $\epsilon$ allows a consistent
perturbative expansion near the fixed point.

 An important feature that emerges clearly in
this approach, and that has been missed in most other treatments
of renormalization group at finite temperature, is the
\emph{anisotropic} scaling between spatial and time directions,
which is manifest in a non-trivial renormalization of the speed of
light.  This is a consequence of the fact that in the Euclidean
formulation at finite temperature time is compatified to $0 \leq
\tau \leq 1/T$, thus space and time or momentum and frequency play
different roles. This results in a novel \emph{dynamical} critical
exponent, $z$ which determines the anisotropic scaling. The
renormalization group leads to scaling in the infrared region in
terms of anomalous dimensions which  can be computed
systematically in the $\epsilon$ expansion. In particular, to
lowest order in $\epsilon$ we find for a scalar theory with
discrete symmetry
$z=1+\frac{\epsilon}{27}+\mathcal{O}(\epsilon^2)$,which describes
a new universality class for dynamical critical phenomena in
quantum field theory. All dynamical aspects, such as the
relaxation rates, and dispersion relations depend on this critical
exponent, while the static aspects are completely described by the
usual critical exponents.

We provide a renormalization group analysis of the quasiparticle
properties near the critical point, such as their dispersion
relation and width, complemented with an explicit evaluation to
lowest order in the $\epsilon$ expansion. The main results
obtained  in this article are the following: at $T=T_c$  we find
that the dispersion relation and width  of quasiparticles of
momentum $p$ is $\omega_p \sim p^{z}$ and $\Gamma_p \sim (z-1) \,
\omega_p$ respectively with a vanishing group velocity of
quasiparticles in the long wavelength limit highlighting the
collective nature of the quasiparticle excitations. For $T>T_c$
but $|T-T_c| \ll T$ we find $\omega_p \sim
\xi^{-z}\left[1+(p\,\xi)^{2z}\right]^{\frac{1}{2}}$ and $\Gamma_p
\sim (z-1)~\omega_p~
 \frac{(p\,\xi)^{2z}}{1+(p\,\xi)^{2z}}$ with $\xi$ the
 finite temperature correlation length $ \xi \propto
 |T-T_c|^{-\nu}$. In the case of $O(N)$ symmetry we find to lowest order in the epsilon
 expansion that the dynamical exponent $z=1+ \epsilon ~
\frac{N+2}{(N+8)^2}+\mathcal{O}(\epsilon^2)$.

 Critical slowing down emerges near the critical point and in the
 $\epsilon$ expansion $\Gamma_p/\omega_p \ll 1$ confirming the
 quasiparticle picture.

 We discuss some relevant cases of threshold singularities in
 which the usual (Breit-Wigner) parametrization of the quasiparticle propagator
 is not available since the real part of the inverse Green's
 function vanishes at the quasiparticle frequency with an anomalous
 power law.

 \vspace{3mm}

 In section II we introduce the model and discuss the breakdown of
 naive  perturbation theory. In section III we introduce the
 $\epsilon$ expansion and analyze the static case. In section IV
 the renormalization aspects and the anisotropic scaling is
 analyzed in detail. Section V presents the renormalization group
 in the effective, dimensionally reduced theory both at and near
 the critical point. This section contains the bulk of our
 results which are summarized in section VI. Our conclusions and a
 discussion of potential  implications are presented in section
 VII. The high temperature behavior of the relevant diagrams is
 computed in the  appendices.

\section{The theory and the  necessity for resummation}\label{section:model}

The low energy sector of QCD with two massless (up and down)
quarks is conjectured to be in the same universality class as the
$O(4)$ Heisenberg ferromagnet\cite{wilraja} described by the
$O(4)$ linear sigma model. Furthermore, since we are interested in
describing the dynamical aspects associated with critical slowing
down and freeze out of long-wavelength fluctuations just before
the chiral phase transition, we focus on $T \rightarrow T_c^+$.

While our motivation for studying critical slowing down stems from
the experimental program in ultrarelativistic heavy ion
collisions, the questions are of a fundamental nature.

To understand the dynamical aspects near the critical point, we
focus on the simpler case of a single scalar field theory, and we
will recover the case of $O(N)$ symmetry at the end of the
discussion. We thus focus on the theory described by the
Lagrangian density \begin{equation} \mathcal{L} = \frac{1}{2}
(\partial_{\mu}\Phi)^2 + \frac{1}{2} m^2_0\Phi^2 -
\frac{\lambda_0}{4!}\Phi^4 \label{barelagrangian} \end{equation}
\noindent where the subscripts in the mass and coupling refer to
bare quantities. The case of $N$ components in the unbroken phase
$T\geq T_c$ differs from the single scalar field by combinatoric
factors that change the critical exponents quantitatively. These
factors will be  at the end of the calculation to obtain an
estimate of the critical exponents for the $O(N)$ theory in
section \ref{sec:summary}.

We are interested in obtaining the relaxation properties of long
wavelength excitations  near the critical temperature, which in
this scalar theory $T_c \sim |m_R|/\sqrt{\lambda_R}$ with
$m_R~;~\lambda_R$ being the renormalized mass and coupling. Thus
the regime of interest for this work is $p,~\omega \ll T \sim T_c$
with $p,~\omega$ being the momentum and frequency of the long
wavelength excitation. As it will become clear below it is
convenient to work in the Matsubara representation of finite
temperature field theory, which is more amenable to the
implementation of the (Euclidean) renormalization group.

In the Matsubara formulation Euclidean time $\tau$ is compactified
in the interval $0\leq \tau \leq \beta=1/T$ whereas space is
infinite, bosonic fields are periodic in Euclidean time and can be
expanded as\cite{kapusta,lebellac,das} \begin{eqnarray}
\Phi(\vec{x},\tau) & = & \frac{1}{\sqrt{\beta V}}
\sum_{n=-\infty}^{\infty}\int \dbarp~ \phi(\vec{p},\omega_n)~
e^{-i\omega_n\tau+i\vec{p}\cdot \vec{x}} \label{FT}\\
\omega_n & = & 2\pi~nT ~~;~~ n=0,\pm 1,\pm 2 \cdots
\label{matsufrec} \end{eqnarray} Thus we see that while the
spatial momentum is a continuum variable, the Matsubara
frequencies are discrete as a consequence of the compactification
of Euclidean time. This feature of Euclidean field theory at
finite temperature will be seen to lead to {\em anisotropic}
rescaling between space and time and therefore, as it will be
clear below, new {\em dynamical} critical exponents. Anticipating
anisotropic rescaling, we then introduce the bare  speed of
propagation $v_0$ of excitations in the medium by writing the
Euclidean Lagrangian in the form \begin{equation} \mathcal{L}_E=
\frac{1}{2} \frac{(\partial_{\tau} \Phi)^2}{v^2_0}+ \frac{1}{2}
(\triangledown \Phi)^2+\frac{1}{2} M^2(T)~\Phi^2 +
\frac{\lambda_R}{4!}~\Phi^4 +\frac{1}{2} \delta m^2(T)~ \Phi^2+
\frac{\delta \lambda}{4!} ~ \Phi^4
\label{lagrangianeuclidean}\end{equation} \noindent where we have
introduced the effective renormalized \emph{temperature dependent}
mass $M(T)$ and the counterterms, in particular the mass
counterterm $\delta m^2(T)=-M^2(T)-m^2_0$ is adjusted order by
order in perturbation theory so that the inverse two point
function obeys \begin{equation} \Gamma^{(2)}(\vec{p}=0;\omega_n=0)
= M^2(T) \label{massofT} \end{equation} The critical point is
defined with (the inverse susceptibility) $M^2(T)\equiv 0$. We
will begin our study by focusing our attention on the critical
theory for which $M(T)=0$. We will later  consider the theory near
the critical point but in a regime in which $M(T)\ll T\sim T_c$.
Thus, the general regime to be studied is $\vec{p},\omega,M(T) \ll
T \sim T_c$.

\subsection{Infrared behavior of the critical theory: static limit in three space dimensions}

In order to highlight the nature of the infrared behavior when
$\vec{p},\omega \ll T$ we focus first on the critical theory in
the static limit, when the Matsubara frequencies of all external
legs in the n-point functions vanish. For the purpose of
understanding the nature of the infrared physics in the static
limit we will set $v_0=1$ in this and next section, and we will
recover this variable, the speed of light, when we study the
dynamics in section \ref{section:dynamics}.

\subsubsection{The scattering amplitude in $D=3$}

 We consider first the $2 \rightarrow 2$
scattering amplitude, or four point function, to one-loop order in
three spatial dimensions. The full expression is given by
\begin{eqnarray} \Gamma^{(4)}({\vec p_1},s_1,{\vec p_2},s_2,{\vec
p_3},s_3,{\vec p_4},s_4) &=& -  \lambda_0+ \lambda^2_0 \left[
H({\vec p_1}+{\vec p_2},s_1+s_2) + H({\vec p_1}+{\vec
p_3},s_1+s_3) \right. \cr \cr &+& \left. H({\vec p_1}+{\vec
p_4},s_1+s_4) \right] +{\cal O}(\lambda^3_0) \; ,
\label{oneloopscattamp} \end{eqnarray} \noindent where $ s_i = 2
\pi \; T\; m_i \; , \, 1 \leq i \leq 4 $ and $ m_i \in {\cal Z} $,
\begin{equation} \label{unlazo} H(p,s) = \frac{T}2 \sum_{n
\epsilon {\cal Z}} \int \dbarq \frac{1}{\left[ q^2 + (2 \pi \; T
\; n)^2 \right] \left[ ({\vec q}+{\vec p})^2 +(2 \pi \; T)^2 \,
(n+m)^2 \right] } \; , \end{equation} \noindent $ s \equiv 2 \pi T
\, m $. Since the external momentum $ p \ll T$ it is clear from
the above expression that the dominant infrared behavior of
$H(p,0)$ is determined by the zero Matsubara frequency in the sum.
As it will be explicitly shown below the contribution from the
non-zero Matsubara frequencies will introduce a renormalization of
the bare coupling which in the limit $T\gg p$ is independent of
the external momentum (this will be seen explicitly in the next
section). Keeping only the zero internal Matsubara frequency and
carrying out the three dimensional integral explicity we find
\begin{equation} H_{ir}(p,0) = \frac{T}{16 p} \label{HIR} \end{equation}
Thus, defining the effective coupling constant at the symmetric
point $\vec{p}_i=\bar{P}_{i}$ where
\begin{equation}\label{sympoint} \bar{P}_i \cdot \bar{P}_j = (4 \;
\delta_{ij} - 1 ) \; {\mu^2 \over 4 } \end{equation} \noindent  in
the static limit one finds that in the infrared limit $\mu/T \ll
1$
\begin{equation} \lambda_{eff}(\mu) = \lambda_0\left[1-\frac{3
\lambda_0 T}{16 \mu} \right]+ \mathcal{O}(\lambda^3_0)
\label{lambdaeff} \end{equation} Two important features transpire
from this expression:~i) the factor $T/\mu$ can be explained by
dimensional arguments: in the Matsubara formulation for each loop
there is a factor $T$ from the sum over internal Matsubara
frequencies. The infrared behavior for $\mu \ll T$ is obtained by
considering only the zero internal Matsubara frequency in the
loop. This integral has only one scale and since $\lambda$ is
dimensionless in three spatial dimensions the one-loop
contribution must be proportional to $T/\mu$. A similar argument
shows that for a diagram with m internal loops, and transferred
momentum scale $\mu$ there will be a power $T^m$ from the
Matsubara sums, the infrared behavior is obtained by the
contribution with \emph{all} the internal Matsubara frequencies
equal to zero, which by dimensional power counting must be of the
form $(T/\mu)^m$. Therefore a diagram with m-internal loops will
contribute to the scattering amplitude with $\lambda (\lambda
T/\mu)^m$. In taking only the zero internal Matsubara frequency we
are assuming that the internal loop momenta are cutoff at a scale
$<T$.

 Thus, at the critical point the most
important infrared behavior is that of the \emph{dimensionally
reduced} three dimensional theory\cite{dimred1,dimred2}. The
reason for this dimensional reduction is clear: at finite
temperature $T$ the Euclidean time is compactified to a cylinder
of radius $L=1/T$ for transferred momenta $\mu$ the spatial
resolution is on distances $d \sim 1/\mu$. Therefore for $\mu \ll
T \rightarrow d \gg L$ thus the compactification radius is
effectively zero insofar as the long distance (infrared) physics
is concerned.

 We
will study below the contribution from the non-zero Matsubara
frequencies.  ii) for a transferred momentum scale $\mu$
perturbation theory breaks down for $\mu \ll \lambda T$ since the
contribution from higher orders is of the form $\lambda (\lambda
T/\mu)^m$. This suggests that a resummation scheme is needed to
study the infrared limit. This situation is similar to that in
critical phenomena where infrared divergences must be summed and
the renormalization group provides a consistent and systematic
resummation procedure. We can obtain a hint of how to implement
the renormalization group in finite temperature field theory in
the limit when $T$ is much larger than any other scale (masses,
momenta and frequency) by realizing that, from the argument
presented above, the  perturbative expansion is actually in terms
of the dimensionless coupling $g_0=\lambda T/\mu$. Therefore from
(\ref{lambdaeff}) we can write \begin{equation} \label{geff}
g_{eff}(\mu)= g_0\left[1-\frac{3}{16}g_0\right] +
\mathcal{O}(g^3_0) \end{equation} We can improve the scattering
amplitude via the renormalization group by considering the
RG-$\beta$ function
\begin{equation}\label{betag}
\beta_g = \mu\frac{\partial g_{eff}(\mu)}{\partial
\mu}\left|_{\lambda_0,T}\right. = -g_{eff}+
\frac{3}{8}g^2_{eff}+\mathcal{O}(g^3_{eff}) \end{equation}
 The first term (with the minus sign) just displays the scaling
 dimension (for fixed $\lambda T$) of the effective coupling, that
 this dimension is $-1$ is a consequence of the dimensional
 reduction since $\lambda T$ is the effective dimensionful
 coupling of the three dimensional theory.

We thus see that the renormalization group improved coupling runs
to the infrared fixed point \begin{equation}\label{fixedpoint}
g^*= \frac{8}{3} \end{equation} \noindent as the momentum scale
$\mu\rightarrow 0$. Comparing with the renormalization group beta
function of critical
phenomena\cite{ma,amit,cardy,zj,wilson,dombgreen} we see that this
is the Wilson-Fischer fixed point in three dimensions, again
revealing the dimensional reduction of the low energy theory. The
resummation of the effective coupling and the fixed point
structure can also be understood in the large N
limit\cite{relaxationI}. As described in\cite{relaxationI} the
large $N$ limit can be obtained by replacing the interaction in
the Lagrangian density for \cite{footnote2}
\begin{equation}\label{largeN} \mathcal{L}_{int} =
\frac{\bar{\lambda}}{2N} \left( \vec{\Phi}\cdot \vec{\Phi}
\right)^2 \end{equation} \noindent with $\vec{\Phi}=(\phi_1,\cdots
\phi_N)$, and the form of the quartic coupling has been chosen for
consistency with the notation of reference\cite{relaxationI}. The
leading order in the large $N$ limit for the scattering amplitude
is obtained by summing the geometric series of one-loop bubbles in
the s-channel (only this channel out of the three contribute to
leading order in the large N limit), each one proportional to $N$
which is the number of fields in the loop. As a result one finds
that the effective scattering amplitude at a momentum transfer
$\mu$ is given by\cite{relaxationI}
\begin{equation}\label{scatamplargeN} \bar{\lambda}_{eff}(\mu) =
\frac{\bar{\lambda}}{1+\frac{\bar{\lambda}T}{4\mu}} \end{equation}
Thus, introducing the dimensionless effective coupling
$\bar{g}_{eff}(\mu)=\bar{\lambda}_{eff}(\mu)(T/\mu)$ one finds
that \begin{equation}\label{fixedpointlargeN} \lim_{\mu\rightarrow
0} \bar{g}_{eff}(\mu)=4 \end{equation} \noindent  i.e, the
effective coupling constant goes to the three dimensional fixed
point\cite{relaxationI}.

\subsubsection{The two point function in $D=3$:} The two point function in the
static limit is given by \begin{equation}\label{defG2}
\Gamma^{(2)}(p,0) = p^2 + \delta m^2(T) - \Sigma(p,0) +{\cal
O}(\lambda^3)  \end{equation} where $ \Sigma(p,0) $ stands for the
two loop sunset diagram at zero external Matsubara frequency and
the counterterm $\delta m^2(T)$ will cancel the momentum
independent, but temperature dependent parts of the self-energy.
The two loop self energy for external momentum $\vec{p}$ and
Matsubara frequency $\omega_m=2\pi T m$ is given by
\begin{equation}\label{SigmaTfull} \Sigma(p,\omega_m) =
\frac{\lambda^2 T^2} {6} \sum_{l,j \epsilon {\cal Z}} \int
\frac{d^3q }{ (2\pi)^3 } \;\frac{d^3k}{(2\pi)^3 } \frac{1}{ [ q^2
+ \omega^2_l] [ k^2 + \omega^2_j] \left[ ({\vec p}+{\vec k}+{\vec
q})^2 + (\omega_l+\omega_j+\omega_m)^2 \right] } \end{equation}
\noindent with $\omega_j = 2\pi jT$.  The static limit is obtained
by setting $\omega_m=0$ ($m=0$). In this limit the dominant
contribution in the infrared for $T\gg p$ arises from the term
$l=j=0$ in the sum. The terms $l\neq 0~;~j\neq0$ for which we can
take $p=0$ (since $p\ll T$) will be cancelled by the counterterm.
A straightforward calculation leads to
\begin{equation}\label{defG2IF} \Gamma^{(2)}(p,0)|_{IR} = p^2
\left [1+
 \frac{\lambda^2 T^2}{12 (4\pi)^2 p^2} \ln\left(\frac{p^2}{\mu^2}\right)\right] +{\cal O}(\lambda^3)
 \end{equation}
\noindent where $\mu$ is a renormalization scale. This expression
clearly reveals the effective coupling $\lambda T/p$ which becomes
very large in the limit $p\ll \lambda T$. Clearly we need to
implement a resummation scheme that will effectively replace the
bare dimensionless coupling constant by the effective coupling
that goes to a fixed point in the long-wavelength limit, and also
ensure that at this fixed point the effective coupling is
\emph{small} so that perturbation theory near this fixed point is
reliable. This is precisely what the renormalization group
combined with the $\epsilon$-expansion achieves in critical
phenomena\cite{ma,amit,cardy,zj,wilson,dombgreen}.

\subsection{Dynamics in $D=3$}

The two-loop contribution to the self-energy for $\omega \neq 0$
is obtained from a dispersive representation of the self-energy in
terms of the spectral density \begin{equation}\label{dispersive}
\Sigma(p,\omega) = \int \frac{d\nu}{\pi}
\frac{\rho(p,\nu)}{\nu-\omega-i0^+} \end{equation} The spectral
density $\rho(p,\nu)$ has been obtained in
reference\cite{relaxationI} in the high temperature
limit\cite{footnote3} in $D=3$. Using the expression given
in\cite{relaxationI} for the spectral density at two loops in the
high temperature limit, and after some lengthy but straightforward
algebra, we find
\begin{equation}\label{specdens} \rho(p,\nu) = \frac{\pi}{12}
\left(\frac{\lambda T}{4\pi} \right)^2 \textmd{sign}(\nu)
\left[\Theta(|\nu|-p)+\frac{|\nu|}{p}\Theta(p-|\nu|) \right]
\end{equation} Carrying out the dispersive integral
(\ref{dispersive}) and subtracting off the terms that are
independent of $p,\omega$ that are absorbed by the counterterm we
find
\begin{equation}\label{twoloopselfenergy} \Sigma(p,\omega) = -\frac{1}{12}
\left(\frac{\lambda T}{4\pi} \right)^2 \left[
\ln\left(\frac{p^2-(\omega+i0^+)^2}{\mu^2} \right)-
\frac{\omega}{p}\ln\left(\frac{\omega+i0^+-p}{\omega+i0^++p}
\right) \right] \end{equation} Clearly, the static limit $\omega
\rightarrow 0$ of the self-energy coincides with (\ref{defG2IF}).
The two point function is therefore given by
\begin{equation}\label{Gamma2pw} \Gamma^{(2)}(p,\omega) =
p^2-\omega^2 +\frac{1}{12} \left(\frac{\lambda T}{4\pi} \right)^2
\left[ \ln\left|\frac{p^2-\omega^2}{\mu^2}\right| -
\frac{\omega}{p}\ln\left|\frac{\omega-p}{\omega+p}\right|
\right]-i\rho(p,\omega) \end{equation} \noindent with
$\rho(p,\omega)$ given by (\ref{specdens}).

There are several features of this expression that are noteworthy:
\begin{itemize}
\item{ It is
clear that for $\lambda T \gg p;\omega$ the two loop contribution
is much larger than the tree level term $p^2-\omega^2$, this
already signals the breakdown of perturbation theory in the high
temperature regime when $\lambda T \gg p;\omega$ in the dynamical
case.  }
\item{Consider the \emph{real part} of the two point function as
$\omega \rightarrow p$ i.e, near  the mass shell.
 \begin{equation}\label{onshell}
 \textit{Re}\Gamma^{(2)}(p, \omega \approx p) \approx 2p^2 \left\{
 \left(1-\frac{\omega}{p}\right)\left[1+\frac{1}{24}
\left(\frac{\lambda T}{4\pi p} \right)^2
\ln\left(\frac{|\omega-p|}{\mu} \right) \right]+ \frac{1}{12}
\left(\frac{\lambda T}{4\pi p} \right)^2 \ln\left(\frac{2p}{\mu}
\right) \right\}\end{equation} This expression reveals that
$\omega =p$ is \emph{not} the position of the mass shell of the
(quasi)-particle. The coefficient of $(1-\omega/p)$ hints at wave
function renormalization but the fact that the two point function
\emph{does not} vanish at this point prevents such identification.
Furthermore, we see that the term that does not vanish at
$\omega=p$ hints at a momentum dependent shift of the position of
the pole, i.e, a correction to the dispersion relation. However
for $\lambda T/p \gg 1$ both contributions are non-perturbatively
large and the analysis is untrustworthy. }

\item{ Now consider the \emph{width} of the quasiparticle,

\begin{equation}\label{twoloopwidth} \gamma_p =
-\frac{\textit{Im}\Sigma(p,\omega=p)}{2p}= \frac{\pi p}{12}
\left(\frac{\lambda T}{4\pi p} \right)^2 \end{equation} \noindent
so that $\gamma_p/p \gg 1$ for $\lambda T/p \gg 1$. This
 signals the breakdown of the quasiparticle picture.

 A similar
analysis reveals   the breakdown of perturbation theory away, but
near the critical point with $|T-T_c|\ll T$. The imaginary part of
the two-loops self-energy at $\vec{p}=0$ and in terms of the
temperature dependent mass $m_R(T)$ can be obtained
straighforwardly in \emph{three spatial
dimensions}\cite{relaxationI,classical} in the limit $T\gg m(T)$.
It is found to be\cite{relaxationI,classical}
 \begin{equation}
\mathrm{Im}\Sigma^{(2)}(\vec{p}=0,\omega=m_R(T)) \propto \lambda^2
T^2 \label{imagi} \end{equation} \noindent consequently, at
two-loops order the width of the zero momentum quasiparticle in
three spatial dimensions is given by\cite{relaxationI,classical}
\begin{equation} \label{gammaaway} \Gamma \propto \frac{\lambda^2
T^2}{m_R(T)} \gg m_R(T) \end{equation} }

\item{This behavior is different from that of gauge theories at high temperature where
low order fermion or gauge boson loops are infrared safe and
determined by the hard thermal loop contributions\cite{brapis}.
This is so because for fermions there is no zero Matsubara
frequency while in the case of gauge bosons, the vertices are
momentum dependent. While the second term inside the bracket in
expression (\ref{Gamma2pw}) is determined by Landau damping is
infrared finite and is similar to the leading contribution in the
hard thermal loop program\cite{brapis,lebellac}, the first term
arises from the three particle cut. The dependence of this term on
the renormalization scale $\mu$ arises from the subtraction of the
mass term at the critical point and reveals the infrared behavior.
Furthermore, in the hard thermal loop
program\cite{brapis,lebellac} one finds thermal masses of order
$gT$ with $g$ the gauge coupling, and widths of order $g^2 T$ (up
to logarithms) so that $\Gamma_p/\omega_p \ll 1$ in the weak
coupling limit, while in the scalar theory under consideration
(\ref{gammaaway}) suggests that $\Gamma_p/\omega_p \gg 1$ in naive
perturbation theory. }

\item{ We note at this stage that the high temperature limit
$p,\omega \ll T$  of the self-energy calculated from the spectral
representation (\ref{dispersive}) can be \emph{directly} obtained
by computing $\Sigma(p,\omega_m)$ in the Matsubara representation
given by (\ref{SigmaTfull}) by setting the internal Matsubara
frequencies $\omega_l=\omega_j =0$ and analytically continuing
$\omega_m \rightarrow -i\omega+0^+$.

We highlight this observation since it will be the basis of
further analysis in what follows: \emph{the high temperature limit
of the self energy $p,\omega \ll T$ can be obtained by setting the
internal Matsubara frequencies to zero and analytically continuing
in the external Matsubara frequency}, i.e,
\begin{equation}\label{zerointmatsu}
 \left.\Sigma(p,\omega)\right|_{p,\omega \ll T} \equiv
\left.{{\lambda^2 T^2 \over 6} \int {d^3q \over (2\pi)^3 } \;{d^3k
\over (2\pi)^3 } \frac{1}{ q^2
 k^2  \left[ ({\vec p}+{\vec k}+{\vec q})^2 +
\omega^2_m\right]}}\right|_{\omega_m\rightarrow -i\omega+0^+}
\end{equation} We provide one and two loops examples of this
statement in appendices \ref{sec:appendix1} and
\ref{sec:appendix2} and formal proof of this statement to one loop
order in appendix \ref{sec:formalproof}.

The result for the width of the quasiparticle at two loops was
anticipated in references\cite{relaxationI,classical}. This width
is purely \emph{classical} since the product $\lambda T$ is
independent of $\hbar$\cite{relaxationI}. This result for the
damping rate of long wavelength quasiparticles in the critical
theory is in striking contrast with that for $T>>T_c$  which has
been studied in detail in\cite{parwani,phi4plasmon}. For $T>>T_c$
the thermal mass is $m_{th} \propto \sqrt{\lambda}T$\cite{parwani}
while the two loop contribution to the imaginary part of the self
energy for $T
>>T_c,p$ is still $\propto \lambda^2 T^2$. Thus, the damping rate of long wavelength excitations
is $\gamma \propto \lambda^{3/2} T \ll \lambda^{1/2}T$ in the weak
coupling limit. Therefore for $T>>T_c$ long wavelength excitations
are true weakly coupled quasiparticles with narrow widths. }
\end{itemize}

\section{The $\epsilon$-expansion: static case}

The analysis of the previous section points out that naive
perturbation theory at finite temperature breaks down at the
critical point for momenta $\ll \lambda T$. The reason for this
breakdown, as revealed by the analysis of the previous section, is
the following: in four space-time dimensions the quartic coupling
is dimensionless, however each loop  diagram in the perturbative
expansion  has a factor $T$ from the sum over the Matsubara
frequencies. After performing the renormalization of the mass
including the finite temperature corrections and setting the
theory at (or near) the critical point, the effective expansion
parameter for long-wavelength correlation functions is $\lambda T$
which has dimensions of momentum. If a given diagram has a
momentum transfer scale $\mu$ the effective \emph{dimensionless}
expansion parameter is therefore $\lambda T/\mu$ which becomes
very large for $\mu \ll \lambda T$, i.e, the effective coupling is
strongly relevant in the infrared. The analysis based on the RG
beta function (\ref{betag},\ref{fixedpoint}) suggests that the
effective coupling $g=\lambda T/\mu$ is driven to the three
dimensional (Wilson-Fischer) fixed point in the infrared,
obviously a consequence of the dimensional reduction in the high
temperature limit. This is confirmed by the large N resummation of
the scattering amplitude
(\ref{scatamplargeN},\ref{fixedpointlargeN}). \emph{If} the value
of the coupling at the fixed point is $\ll 1$ then a perturbative
expansion \emph{near the fixed point} would be reliable, however
the value of the coupling at the fixed point is $g^* \sim
\mathcal{O}(1)$ which of course is a consequence of the fact that
for fixed $\lambda T$ the effective coupling scales with dimension
of inverse momentum in the infrared. This situation is the same as
in critical phenomena for theories that are superrenormalizable,
in which the infrared divergences are severe.

The remedy in critical phenomena is to study the perturbative
series via the $\epsilon$-expansion wherein the value of the
coupling at the fixed point is
$\mathcal{O}(\epsilon)$\cite{ma,amit,cardy,zj,dombgreen} and sum
the perturbative series via the renormalization group. We now
implement this program in the high temperature limit.

In \emph{five} space-time dimensions the quartic coupling
$\lambda$ has canonical dimension of inverse momentum, therefore
the product $\lambda T$ that occurs in the perturbative expansion
in the dimensionally reduced low energy theory is
\emph{dimensionless}. Then in a perturbative expansion at (or very
near) the critical point we expect that infrared divergences will
be manifest in the form of logarithms of the momentum scale in the
loop. This implies that the effective coupling is \emph{marginal}.
Considering the theory in $4-\epsilon$ spatial dimensions and one
Euclidean (compactified) time dimension the effective coupling of
the dimensionally reduced theory, $\lambda T$,  has dimensions of
$\mu^{\epsilon}$ with $\mu$ being a momentum scale. Therefore the
effective dimensionless coupling for diagrams with a transferred
momentum scale $\mu$ is $g(\mu)= \lambda T \mu^{-\epsilon}$. Thus,
for fixed $T$ the scaling dimension of this effective coupling is
$-\epsilon$, hence we expect a non-trivial fixed point at which
the coupling $g^*\sim \mathcal{O}(\epsilon)$.

Therefore for $\epsilon \ll 1$ we can perform a systematic
perturbative expansion near the fixed point. This is the spirit of
the $\epsilon$ expansion in critical phenomena which, when
combined with a resummation of the perturbative series via the
renormalization group has provided a spectacular quantitative and
qualitative understanding of critical
phenomena\cite{wilson,ma,amit,cardy,zj,dombgreen}.

While dimensional regularization and the $\epsilon$ expansion have
been used to study the dimensionally reduced high temperature
theory insofar as thermodynamic quantities is concerned, i.e,
static phenomena\cite{irla,dimred1,dimred2}, we emphasize that our
focus is to study \emph{dynamics} at and near the critical point,
which is fundamentally different from the studies of static
phenomena in these references.

As a prelude to the study of the dynamics, we now revisit the
scattering amplitude at one loop level and the self-energy at two
loops level  in $4-\epsilon$ \emph{spatial} dimensions at high
temperature in the static limit. The one loop self-energy is
momentum independent and is absorbed in the definition of the
thermal mass\cite{parwani} which is set to zero at the critical
point. The are two main purposes of this exercise: the first is to
quantify the role of the higher Matsubara modes and secondly to
obtain a guide for the infrared running of the coupling constant.

\subsection{Scattering amplitude:}

The one loop contribution in the static limit, i.e, when the
external Matsubara frequencies are zero is given by
\begin{equation} \label{unlazoepsi} H(p,0) = \frac{T}{2}  \sum_{n
\epsilon {\cal Z}} \int \frac{d^dq}{(4\pi)^d}{1 \over \left[ q^2 +
(2 \pi \; T \; n)^2 \right] \left[ ({\vec q}+{\vec p})^2 +(2 \pi
\; T\;n)^2 \right] } ~~;~~ d= 4-\epsilon \; , \end{equation} In
the high temperature limit the nonzero Matsubara terms give
subdominant contributions. This property can be argued in
different manners. For example, the $ l \neq 0 $ terms in
eq.(\ref{unlazoepsi}) can be interpreted as Feynman diagrams in
$d= 4 - \epsilon $ dimensions with mass $  (2 \pi \; T \; l)^2 $.
Such contributions are negligible for $ p \ll T
$\cite{laplata,apple}.

We find for high temperatures (see Appendix \ref{sec:appendix1} ),
\begin{equation} H(p,0) \buildrel{T \gg p,s}\over = H_{asi}(p,0) -
T^{1-\epsilon}\frac{\Gamma\left(1 + \frac{\epsilon}2 \right) \;
\zeta(2+\epsilon)}{ 192 \; \pi^{4+\frac{\epsilon}{2}}}
\frac{p^2}{T^2} \left[ 1 + {\cal
O}\left(\frac{p^2}{T^2}\right)\right] \; , \end{equation} where $
H_{asi}(p,) $ stands for the $ l=0$ contribution to the sum
(\ref{unlazoepsi} ) {\bf plus} the dominant high temperature limit
of the sum over the $ l \neq 0 $ terms, which is obtained by
setting $p=0$. Separating the $l=0$ mode and setting $p=0$ in the
contribution from the sum over $l\neq 0$  we find
\begin{equation}\label{epsiHiT} H_{asi}(p,0)=\frac{T p^{-\epsilon}
}{2(4\pi)^{2-\frac{\epsilon}{2}}}~
\Gamma\left(\frac{\epsilon}{2}\right)\left[\frac{\Gamma^2(1-\frac{\epsilon}{2})}{\Gamma(2-\epsilon)}
+  2 \left(\frac{4\pi^2 T^2}{p^2} \right)^{-\frac{\epsilon}{2}}
\zeta(\epsilon) \right]\end{equation} \noindent with $\zeta$
Riemann's zeta function which has the following
properties\cite{gr}
\begin{equation}\label{riemann} \zeta(0)= -\frac{1}{2} ~~;~~ \lim_{\epsilon
\rightarrow 1}\zeta(\epsilon) = \frac{1}{\epsilon-1}+ \gamma
\end{equation} \noindent where $  \gamma = 0.577216\ldots $ stands
for the Euler-Mascheroni constant.

Near four space dimensions $\epsilon \rightarrow 0$ we find
\begin{equation}\label{dim4spacedim} H_{asi}(p,0) = -  { T \over 2 (4\pi)^2 }
\left[ \log\left({ p^2 \over 4 \, T^2}\right) - 2 \right] + {\cal
O}(\epsilon)\; . \end{equation}

There is \emph{no pole} in $\epsilon$ and the argument of the
logarithm reveals that $T$ acts as an ultraviolet cutoff. The
reason that there is no pole in epsilon as
 $\epsilon \rightarrow 0$ is that the poles in $\epsilon$ should
 be independent of temperature and should be those of the zero
 temperature theory. However, in dimensional regularization one loop integrals
 have no poles in odd space-time dimensions\cite{dimreg}. On
the other hand near
 three space dimensions $\epsilon \rightarrow 1$ we find
 \begin{equation}
H_{asi}(p,0) = {T \over 16 \,  p}\left[1  -{1 \over \pi^2}
\frac{p}{T}\ln{T \over \mu} \right]+ {1 \over (4\pi)^2 \;
(\epsilon - 1)} \; .
\end{equation}
where the pole term at $ \epsilon = 1 $ corresponds to the usual
coupling constant renormalization. This divergent term is
temperature-independent, as expected, the $\ln(T)$ is reminiscent
of an upper momentum cutoff for the high temperature limit. The
first term is precisely what we obtained in equation (\ref{HIR})
from setting the internal Matsubara frequency to zero, i.e, the
result of the dimensionally reduced theory.  After subtracting the
pole near three space dimensions, the first term gives the leading
infrared contribution in the limit $T/p \gg 1$ whereas the
logarithm is subleading.

This expression coincides with that given in
references\cite{dimred1,dimred2}. In these references the four
dimensional high temperature static theory was studied and a
systematic analysis of Feynman diagrams in the dimensionally
reduced theory (three dimensions) was performed. The $\epsilon-1$
in our expressions should be mapped onto the $2\epsilon$ for
comparison with the results in these references.

For $\epsilon >0$ we can neglect the terms of the form
$(T/p)^{-\epsilon}$ in eqn. (\ref{epsiHiT}) in the limit $T/p\gg
1$. And for $1\gg \epsilon >0$ we find that the static scattering
amplitude at a symmetric point (\ref{sympoint}) in the limit in
which the temperature is much larger than the external momentum
scales is given by \begin{equation}\label{gamma4eps}
\Gamma^{(4)}(p_i=\bar{P}_i,0)=-\lambda_{eff}(\mu,T)=
-\lambda\left[1 - \frac{3}{(4\pi)^{\frac{d}{2}}}\frac{\lambda T
\mu^{-\epsilon}}{\epsilon} \right] \end{equation} \noindent the
factor $T$ arising from the Matsubara sum is such that $\lambda T$
has dimensions of $(\textmd{momemtum})^{\epsilon}$ so that in
$5-\epsilon$ space time dimensions $\lambda T \mu^{-\epsilon}$ is
\emph{dimensionless}. Thus, introducing the \emph{dimensionless}
renormalized coupling \begin{equation}\label{dimlesscoupren}
g_R(\mu)= \frac{\lambda_{eff}(\mu,T)T
\mu^{-\epsilon}}{(4\pi)^{\frac{d}{2}}} ~;~
d=4-\epsilon\end{equation} \noindent we find
\begin{equation}\label{betafunc}
\beta_g = \left.\mu \frac{\partial g}{\partial
\mu}\right|_{\lambda,T} = -\epsilon g+3 g^2
+\mathcal{O}(g^3)\end{equation} Therefore this effective coupling
in the infrared limit is driven to a non-trivial fixed point
\begin{equation}\label{fpcoup} g^* = \frac{\epsilon}{3}\end{equation} Hence
for $\epsilon \ll 1$ the fixed point theory can be studied
perturbatively. This of course is the basis of the $\epsilon$
expansion in critical
phenomena\cite{ma,wilson,amit,cardy,zj,dombgreen} and will be the
important point upon which our analysis will hinge.

\subsection{Two loops self-energy:}
As mentioned above the one loop contribution to the self energy is
momentum independent and absorbed into the definition of the
thermal mass. The two loop contribution in the static limit in $d$
\emph{spatial} dimensions is given by
\begin{equation}\label{2lupsigma} \Sigma^{(2)}(k,0)=
\frac{\lambda^2 T^2}{6} \sum_{l} \sum_{m} \int\ddbarp \ddbarq
\frac{1}{[q^2+\omega^2_l][p^2+\omega^2_m][(p+q+k)^2+(\omega_l+\omega_m)^2]}
\end{equation} We now introduce two Feynman parameters, separate the $l=m=0$
term from the Matsubara sums and take $T\gg p$ in the sums with
$l,m \neq 0$ to find \begin{equation}\label{sig2hiT}
\Sigma^{(2)}(k,0)=\frac{g^2}{6}
\frac{\Gamma(-1+\epsilon)\Gamma^3(1-\frac{\epsilon}{2})}{\Gamma(3-\frac{3\epsilon}{2})}k^2
\left[\frac{k^2}{\mu^2} \right]^{-\epsilon}+  g^2
T^{2}\left[\frac{T^2}{\mu^2} \right]^{-\epsilon} \mathcal{C}(d)
\end{equation} \noindent with $\mathcal{C}(d)$ only depending on the
dimensionality. The second term $\propto
 T^{4-2\epsilon}$ does not depend on the momentum and is therefore
 cancelled by the mass counterterm which defines the critical
 theory. Therefore for $\epsilon >0$ but small we find
\begin{equation}\label{Gamma}
 \Gamma^{(2)}(k,0)= p^2 -\left[\Sigma(k,0)-\Sigma(0,0)\right]= k^2\left[1+\frac{g^2}{12\epsilon}-\frac{g^2}{12}
 \ln\left(\frac{k^2}{\mu^2}\right) \right]+ \mathcal{O}(g^3)\end{equation}
We introduce the wave function renormalization in dimensional
regularization by the usual relation
\begin{equation}\label{gamma2ren} \Gamma^{(2)}_R(k,0)=
Z_{\phi}~\Gamma^{(2)}(p,0) \end{equation} \noindent and choose
\begin{equation}\label{wavefuncren} Z_{\phi}=
1-\frac{g^2}{12\epsilon}+\mathcal{O}(g^3) \end{equation} Therefore
the renormalized two point function in the static, high
temperature limit is given by \begin{equation}\label{twopointren}
\Gamma^{(2)}_R(k,0)=
k^2\left[1-\frac{g}{12}\ln\left(\frac{k^2}{\mu^2}\right)
\right]+\mathcal{O}(g^3)\end{equation} The infrared behavior is
obtained by resumming the perturbative series via the
renormalization group\cite{amit,cardy,zj,ma,wilson,dombgreen}
which leads to the scaling form of the two point function in the
infrared limit
\begin{equation}\label{RGsummed2} \Gamma^{(2)}_R(k) \propto k^{2-\eta} \end{equation}
\noindent with \begin{equation}\label{eta} \eta=\frac{{g^*}^2}{6}=
\frac{\epsilon^2}{54}+\mathcal{O}(\epsilon^3) \end{equation} This
is the anomalous dimension to lowest order in
$\epsilon$~\cite{ma,wilson,amit,cardy,zj,dombgreen}.

\bigskip

\subsection{The strategy:}
The analysis of the static case above has highlighted several
important features of the infrared behavior near the critical
point, which determines the strategy for studying the dynamical
case: {\bf a:)} the infrared behavior in the limit when $T \gg p$
with $p$ the typical momentum of the Feynman diagram is determined
by the dimensionally reduced theory obtained by setting the
\emph{internal} Matsubara frequencies to zero. {\bf b:)} naive
perturbation theory breaks down in three space dimensions because
the dimensionless coupling is $\lambda T/\mu$ with $\mu$ the
external momentum scale in the Feynman diagram, while a large $N$
or renormalization group resummation suggests a non-trivial
infrared stable fixed point, the coupling at this fixed point is
of $\mathcal{O}(1)$. {\bf c:)} Just as in critical phenomena the
perturbative expansion can be systematically controlled in an
$\epsilon$ expansion around \emph{four spatial dimensions}
corresponding to a theory dimensionally reduced from $5-\epsilon$
space time dimensions. The effective dimensionless coupling of the
dimensionally reduced theory (four dimensional) is $\lambda T$
which is marginal. This combination is independent of $\hbar$-this
can be seen by restoring powers of $\hbar$ $\lambda \rightarrow
\hbar \lambda; T \rightarrow T/\hbar$ - so that the low energy,
dimensionally reduced theory is \emph{classical}. In $4-\epsilon$
spatial dimensions the effective dimensionless coupling $g=\lambda
T \mu^{-\epsilon}$ is driven to the infrared stable Wilson-Fischer
fixed point of $\mathcal{O}(\epsilon)$ by the renormalization
group trajectories. Thus, the strategy to follow becomes clear: we
will now study the dynamics by including the contribution from the
external Matsubara frequency, focusing on the infrared behavior
for $p,\omega \ll T$ near the critical point in a systematic
$\epsilon$ expansion around \emph{four} spatial dimensions.

We note that the theory in $5-\epsilon$ dimension is formally
non-renormalizable in the ultraviolet, however this is irrelevant
for the infrared which is the region of interest here. The
analysis provided above in the limit $\epsilon \rightarrow 0$
clearly shows that near five space-time dimensions there are no
poles in dimensional regularization in one loop diagrams as
expected\cite{laplata}. The potential poles are replaced by
$\ln(T)$. The low-energy theory must be understood with a cutoff
of $\mathcal{O}(T)$ and the dimensionally regularized integrals in
5 space time dimensions clearly display this cutoff in the
argument of logarithms. The long wavelength $\mu/T\ll 1$ an the
$\epsilon \rightarrow 0$ limits do not commute: keeping the
subleading terms in the high temperature limit and taking
$\epsilon \rightarrow 0$ results in that poles in $\epsilon$
actually translate into logarithms of the cutoff $T$, on the other
hand, keeping $\epsilon
>0$ and small the $T/\mu \rightarrow \infty$ limit can be taken and
the subleading high temperature corrections vanish. Clearly it is
the \emph{latter} limit the one that has physical relevance, since
eventually we are interested in studying the infrared behavior of
the physical theory in three space dimensions. Hence in what
follows we consider the long-wavelength limit for $\epsilon >0$
but small and approach the physical dimensionality $\epsilon
\rightarrow 1$ in a consistent $\epsilon$ expansion improved via
the renormalization group. This is the strategy in classical
critical phenomena as well where for $\epsilon >0$ and small the
ultraviolet cutoff can be taken to infinity.

\bigskip

 At this stage it is important to highlight the difference between
the main focus of this work and that in
references\cite{dimred1,dimred2}. The work of
references\cite{dimred1,dimred2} studies the \emph{static} limit
of the dimensionally reduced theory near \emph{three} spatial
dimensions arising from the high temperature limit of a four
dimensional Euclidean theory compactified in the time direction.
In contrast we here focus on studying the \emph{dynamics} in the
limit when $p,\omega \ll T$ which as emphasized by the analysis
above will be studied in an $\epsilon$ expansion in a
dimensionally reduced theory near \emph{four} space dimensions.

The limit of physical interest $\epsilon \rightarrow 1$ must be
studied by improving the perturbative expansion via the
renormalization group\cite{amit,cardy,zj,ma,wilson,dombgreen} and
eventually by other non-perturbative resummation methods, such as
Pade approximants or Borel resummation that will extend the regime
of validity of the $\epsilon$ expansion\cite{zj}.

\section{Dynamics near the critical
point:}\label{section:dynamics} We now turn to the dynamics. Our
main goal is to study the feasibility of a quasiparticle
description of low energy excitations at and near the critical
point. Of particular interest is the dispersion relation as well
as the damping rates of these excitations. This information is
contained in the two point function $\Gamma^{(2)}(p,\omega_n)$
which is the inverse propagator, analytically continued to
$\omega_n \rightarrow -i\omega+0^+$. The region of interest is
$p,\omega \ll T$ and if the theory is (slightly) away from the
critical point $M(T)\ll T$ as well. In principle for a fixed
$\omega_n$ or fixed external Matsubara frequencies in the external
legs of $n-point$ functions, one must perform the sum over the
internal Matsubara first and then take the analytic continuation.
However, as was shown above in detail in the static case, the most
infrared singular contribution arises from setting the internal
Matsubara frequencies to zero. That this is  also the case in the
dynamics can be seen  by considering a diagram with $m-$ internal
lines, rerouting the external Matsubara frequency through one of
the lines. All of the lines are equivalent since rerouting the
external Matsubara frequency corresponds to a shift in one of the
sums. The other $m-1$ contain propagators in which the internal
Matsubara frequency acts as a mass of $\emph{O}(2\pi l~T)$. These
are the superheavy modes in the description of
references\cite{dimred1,dimred2,laplata,apple}. The contribution
that is dominant in the infrared is from the region of loop
momenta $\ll T$ which is largest when the mass of the propagator
is zero, i.e, the zero Matsubara frequency. Keeping non-zero
Matsubara frequencies in any of the $m-1$ legs will lead to
subleading contributions in the limit $p,\omega \ll T$.

Once the internal Matsubara frequencies had been set to zero we
can analytically continue the external Matsubara frequency to a
continuous Euclidean variable $\omega_n\rightarrow s$ to obtain
the Euclidean two point function. The dispersion relation and
damping rate are obtained by further analytical continuation
$s\rightarrow -i\omega+0^+$.

\bigskip

As anticipated in section(\ref{section:model}) because Euclidean
time is compactified and plays a \emph{different} role than the
spatial dimensions, we must consider the anisotropic Lagrangian
density (\ref{lagrangianeuclidean}) which includes the velocity of
light multiplying the derivatives with respect to Euclidean time.
If this velocity of light is simply a constant it can be
reabsorbed into a trivial redefinition of the time variable.
However, as it will become clear below, this velocity of light
acquires a non-trivial renormalization as a consequence of the
anisotropy between space and time directions at finite temperature
and will run with the renormalization group. Thus, the Euclidean
propagator is generalized to \begin{equation}\label{anisotroprop}
G(k,\omega_m) =
\frac{1}{k^2+\frac{\omega^2_m}{v^2_0}}\end{equation}

\subsection{The scattering amplitude}

We begin by studying the scattering amplitude, now as a function
of external momenta and frequencies. The one loop contribution is
determined by the function $H(p,s)$ given by eqn. (\ref{unlazo}),
which for $p,s \ll T$ is given by (see Appendix) \begin{equation}
H(p,s) \buildrel{T \gg p,s}\over = H_{asi}(p,s) - \frac{
\Gamma\left(1 + \frac{\epsilon}2 \right) \; \zeta(2+\epsilon)}{
192 \; \pi^{4+\frac{\epsilon}{2}} \; T^{1 + \epsilon}} \left[ p^2
+ \frac{s^2}{v^2_0}(1-\epsilon) \right] \left[ 1 + {\cal
O}\left(\frac{p^2 , \frac{s^2}{v^2_0}}{T^2}\right)\right] \; ,
\end{equation} where $ H_{asi}(p,s) $ stands for the $ l=0$
contribution to the sum (\ref{unlazo} ) {\bf plus} the dominant
high temperature limit of the sum over the $ l \neq 0 $ terms
which is obtained by setting $p,s=0$
\begin{equation}gin{equation}
H_{asi}(p,s) = \frac{T}{2}  \int \frac{d^d q}{(2\pi)^d } \frac{1}{
q^2 \left[ ({\vec q}+{\vec p})^2 + \frac{s^2}{v^2_0} \right] } +
\frac{\Gamma\left(1 + \frac{\epsilon}2 \right) \; \zeta(\epsilon)}
{ 8 \; \pi^{2+\frac{\epsilon}2} \; \epsilon } \; T^{1 - \epsilon}
\;,
\end{equation}
This integral is computed in the appendix A with the result
\begin{eqnarray} \label{4ptonu} &&H_{asi}(p,s) = - \frac{\Gamma\left(
\frac{\epsilon}
 {2}-1\right)}{2\left( 4 \pi \right)^{2-\frac{\epsilon}{2}}}
 T\left(\frac{s^2}{v^2_0} + p^2\right)^{ - \frac{\epsilon}{2}} \;
F\left(\frac{\epsilon}{2},1-\frac{\epsilon}{2};2-\frac{\epsilon}{2};\frac{p^2}
{p^2 + \frac{s^2}{v^2_0}}\right) + \cr \cr &&+ \;
\frac{\Gamma\left(1 + \frac{\epsilon}{2} \right) \;
\zeta(\epsilon)}{  8 \; \pi^{2+\frac{\epsilon}{2}} \; \epsilon }
\; T^{1 - \epsilon} \; , \end{eqnarray} \noindent where $
F(a,b;c;z) $ stands for the hypergeometric function. For $\epsilon
>0$ and $p,\frac{s}{v_0} \ll T$ we can neglect the second term,
since it is proportional to $T^{1-\epsilon}\ll
T\;(p^2+\frac{s^2}{v^2_0})^{-\frac{\epsilon}{2}}$. We note that
the infrared dominant contribution can be written in the form $T
p^{-\epsilon}\mathcal{F}(\frac{s^2}{v^2 p^2})$ the factor $T$ thus
combines with the coupling $\lambda$ to give the effective
coupling of dimension $\mu^{\epsilon}$ in $d=4-\epsilon$ spatial
dimensions.

For $\epsilon >0$ but small and neglecting the second term in
(\ref{4ptonu}) can be expanded in $\epsilon$ leading to
\begin{equation} \label{dim5pt} \lambda_0 H_{asi}(p,s)  =
\frac{g(\mu)}{2}\left[ \frac{2}{\epsilon} -\left( 1 +
\frac{s^2}{v^2_0 p^2} \right)\ln\left(\frac{s^2
+p^2v^2_0}{\mu^2v^2_0}\right) + \frac{s^2}{v^2_0
p^2}\ln\left(\frac{s^2}{v^2_0 \mu^2}\right) + \ln 4\pi + 2 -
\gamma + {\cal O}(\epsilon) \right] \end{equation} where we
introduced the dimensionless bare coupling
\begin{equation}\label{dimenlessbare} g(\mu) =\frac{\lambda_0 T
\mu^{-\epsilon}}{(4\pi)^{\frac{d}{2}}} \end{equation} We remark
that   one cannot take $\epsilon \rightarrow 0$ in this expression
since in this limit the pole is actually cancelled by the second
term in (\ref{4ptonu}) above. As emphasized above, this expression
must be understood for $\epsilon >0$ but small so that the
contributions of the form $\left(T/s,T/p
\right)^{-\epsilon}\rightarrow 0$ for $T \gg s,p$. Therefore, the
expression above must be understood in the sense that i): $T\gg
s,p$ with fixed $\epsilon >0$ and ii): $\epsilon \ll 1$ and the
resulting expressions have a Laurent expansion for small
$\epsilon$.

\subsection{The self energy at two loops}
Neglecting the contribution from the non-zero Matsubara
frequencies which will be absorbed by the mass counterterm in the
definition of the critical temperature (or $M^2(T)$ away from the
critical point) and also neglecting terms that vanish in the limit
$T\gg p,s$ the dominant contribution in the infrared to the two
loops self-energy is

\begin{eqnarray}\label{self2lupir} \Sigma(p,s) = && \frac{\lambda^2_0
T^2}{6}\int \ddbarq \int \ddbark
\frac{1}{q^2k^2\left[(q+k+p)^2+\frac{s^2}{v^2_0} \right]}
\nonumber \\
=&& \frac{\lambda^2_0
T^2}{6(4\pi)^d}~\frac{\Gamma^2(\frac{d}{2}-1)\Gamma(3-d)}{\Gamma(d-2)}~\int^1_0
dx (1-x)^{d-3}x^{1-\frac{d}{2}}\left[xp^2+\frac{s^2}{v^2_0}
\right]^{d-3} \end{eqnarray} While this expression can be written
in terms of hypergeometric functions, it is more convenient to
expand it in $\epsilon$ with the result that
\begin{equation}\label{gammaeps}
 \Gamma^{(2)}(p,s)=
p^2+\frac{s^2}{v^2_0}+\frac{g^2(\mu)}{6\epsilon}\left[\frac{p^2}{2}+\frac{2}{\epsilon}\frac{s^2}{v^2_0}-
\epsilon~
\frac{p^2}{2}\ln\left(\frac{p^2}{\mu^2}\right)-2\frac{s^2}{v^2_0}\ln\left(\frac{s^2}{v^2_0\mu^2}
\right) \right]+\mathcal{O}(g^2\epsilon^0,g^2\epsilon,g^3)
\end{equation} \noindent with $g(\mu)$ given by eqn.
(\ref{dimenlessbare}).

\subsection{Renormalization}\label{sec:renor}

The forms of the two  and four  point functions immediately
suggests a renormalization scheme akin to the familiar one used in
critical phenomena\cite{amit,cardy,zj,dombgreen} with \emph{one
important difference}: we see from eqn. (\ref{gammaeps}) that the
velocity of light $v_0$ must also be renormalized. The wave
function renormalization is introduced as usual via
\begin{equation}\label{rentwopoint} \Gamma^{(2)}_R(p,s,v_R) =
Z_{\phi}~\Gamma^{(2)}(p,s,v_0) \end{equation} \noindent the
renormalized mass
 as a function of temperature
is defined as

\begin{equation}\label{massren} \Gamma^{(2)}_R(0,0)= M^2(T) \end{equation}

\noindent this definition, however, defines the inverse
susceptibility or correlation length, rather than the pole mass,
the critical theory is defined by $M^2(T)=0$. Coupling constant
and velocity of light renormalization are achieved by
\begin{eqnarray}\label{couprenor} && \lambda_R = Z_{\phi}^2 ~Z_{\lambda}~
\lambda_0 \nonumber \\
&& \Gamma^{(4)}_R(p_i=\bar{P}_i,s=0) = -\lambda_R \end{eqnarray}
\begin{equation}\label{renvelo} v^2_R= v^2_0 \frac{Z_v}{Z_{\phi}} \end{equation} The
renormalization conditions that determine the constants
$Z_{\phi},Z_{\lambda}, Z_v$ are \begin{eqnarray}\label{renconds}
&& \left.\frac{\partial \Gamma^{(2)}_R} {\partial
p^2}\right|_{p^2=\mu^2,\frac{s^2}{v^2_R}=\mu^2}= 1 \nonumber
\\
&&\left.\frac{\partial \Gamma^{(2)}_R}{\partial
s^2}\right|_{p^2=\mu^2,\frac{s^2}{v^2_R}=\mu^2}=
\frac{1}{v^2_R}\nonumber \\
&& \Gamma^{(4)}(p_i=\bar{P}_i;s=0) = -\lambda_R   \end{eqnarray}
Consistently with the $\epsilon$ expansion we choose the
renormalization constants $Z_{\phi},Z_{\lambda}, Z_v$ in the
minimal subtraction scheme to lowest order, since keeping higher
powers of the coupling or $\epsilon$ results in higher order
corrections in the $\epsilon$ expansion.

To lowest order, one loop for the 4-point function and two loops
for the two point function, we find from the results
(\ref{dim5pt}) and (\ref{gammaeps}) \begin{eqnarray} &&
Z_{\lambda}= 1-\frac{3g(\mu)}{\epsilon} \Rightarrow g_R=
g(\mu)-\frac{3g^2(\mu)}{\epsilon} ~~;~~ g_R={\lambda_R \, T \,
\mu^{-\epsilon}\over (4\pi)^{\frac{d}{2}}} \label{cupre} \cr \cr
&& Z_{\phi} = 1-\frac{g^2}{12\epsilon} \label{zphi} \\
&&Z_v = 1- \frac{g^2}{3\epsilon^2} \label{zvel} \end{eqnarray}
Thus, the renormalized two point function reads
\begin{equation}\label{2ptren} \Gamma^{(2)}_R(p,s)=
p^2\left[1-\frac{g^2}{12}\ln\left( \frac{p^2}{\mu^2}\right)
\right]+\frac{s^2}{v^2_R}\left[1- \frac{g^2}{3\epsilon}\ln\left(
\frac{s^2}{v^2_R\mu^2}\right)\right]+\mathcal{O}(g^3,g^2\epsilon)
\end{equation}

\section{The renormalization group:}

Before we embark on the resummation program via the
renormalization group, it is important to highlight two important
features

\begin{itemize}
\item{The contributions that are dominant in the infrared in the limit $T\gg p,s$ correspond
to the terms with  internal Matsubara frequencies equal to zero,
the non-zero Matsubara frequencies give subleading contributions
for $\epsilon >0$. This in turn results in that the dependence on
temperature is solely through the effective coupling $g=\lambda T
\mu^{-\epsilon}$. This can be seen from the fact that each loop
has a factor T from the Matsubara sum over the internal loop
frequencies as well as one power of the coupling constant
$\lambda$, by dimensional reasons the dimensionless coupling is
obtained by multiplying by $\mu^{-\epsilon}$.  }

\item{The velocity of light $v$ always enters in the form $s/v$ since this is the form that enters in the
propagators and the renormalization conditions above.  }

\end{itemize}

\subsection{The critical point:}

 The bare n-point functions are independent of the
renormalization scale $\mu$, and this independence leads to the
renormalization group equations (we now suppress the subscript R
understanding that all quantities are renormalized)
\begin{equation}\label{egr} \left[ \mu {\partial \over
\partial\mu} + \beta_{g}\frac{\partial}{\partial g} +
\beta_v\frac{\partial}{\partial v} - {N \over 2} \; \gamma
\right]\Gamma^{(N)}\left(p_1,\frac{s_1}{v};
p_2,\frac{s_2}{v}\ldots,p_N;\frac{s_N}{v}; g,\mu\right) = 0
\end{equation} \noindent with \begin{eqnarray}
\beta_g & =&  \left.\mu \frac{\partial g}{\partial \mu}\right|_{\lambda_0,T,v_0} \label{betageps} \\
\beta_v & = & \left.\mu \frac{\partial v}{\partial
\mu}\right|_{\lambda_0,T,v_0}\label{betav} \\
\gamma  & = & \left.\mu \frac{\partial \ln Z_{\phi}}{\partial
\mu}\right|_{\lambda_0,T,v_0} \label{gammaexp} \end{eqnarray} To
lowest order we find \begin{eqnarray}  &&\beta_g = -\epsilon~g+ 3
g^2 + \mathcal{O}(g^3,g^2\epsilon)
\label{betag1lup}\\
&&  \beta_v=
\frac{1}{2}\left[\frac{2g^2}{3\epsilon}-\gamma\right]~v+
\mathcal{O}(g^3,g^2\epsilon)
\label{betav2lup}\\
&&\gamma=\frac{g^2}{6}+ \mathcal{O}(g^3,g^2\epsilon)
\label{gamma2lup} \end{eqnarray}

While we can write down the general solution of the RG equation
(\ref{egr}) for an arbitrary $N$-point function, our focus is to
understand the quasiparticle structure which is obtained from
$\Gamma^{(2)}$.

Since $\Gamma^{(2)}$ has dimension two, it follows that

\begin{equation} \label{dimlessgamma} \Gamma^{(2)}(p,\frac{s}{v},g,\mu) =
\mu^2~ \Phi\left(\frac{p}{\mu},\frac{s}{v\mu},g\right)
\end{equation}

\noindent therefore \begin{equation}\label{sca}\Gamma^{(2)}(e^t
p,\frac{e^t s}{v},g,\mu)= e^{2t}~\Gamma^{(2)}(p,\frac{s}{v},g,\mu
e^{-t}) \end{equation} This scaling property then leads to
\begin{equation}\label{dteqn} \left[\frac{\partial}{\partial t}+
\mu \frac{
\partial}{\partial \mu}-2\right]\Gamma^{(2)}(e^t p,\frac{e^t
s}{v},g,\mu)=0 \end{equation} \noindent which combined with the RG
equation (\ref{egr}) leads to the following equation that
determines the scaling properties of the two point function
\begin{equation} \label{dtgamma2} \left[ -
\frac{\partial}{\partial t} + \beta_{g}\frac{\partial}{\partial g}
+ \beta_v\frac{\partial}{\partial v} - ( \gamma -2)
\right]\Gamma^{(2)}\left(e^t p,\frac{e^t s}{v},g,\mu\right) = 0
\end{equation} The solution of this equation is standard\cite{amit,cardy,zj}
\begin{equation}\label{rgsol} \Gamma^{(2)}\left(e^t p,\frac{e^t
s}{v},g,\mu\right)= e^{\int_0^t
dt'\left(2-\gamma(t')\right)}~\Gamma^{(2)}\left(
p,\frac{s}{v(t)},g(t),\mu\right)\end{equation} \noindent with
\begin{eqnarray}\label{running} \frac{\partial g(t)}{\partial t} & = &
\beta_g(g(t))~;~ g(0)=g_R(\mu) \nonumber \\
\frac{\partial v(t)}{\partial t} & = & \beta_v(v(t),g(t))~ ; ~
v(0)=v_R(\mu)\nonumber
\\
\gamma(t)& = & \gamma(g(t),v(t)) \end{eqnarray} As $t\rightarrow
-\infty$ i.e, the momentum and frequency are scaled towards the
infrared we see from  the RG $\beta$ function (\ref{betag1lup})
that coupling is driven to its fixed point
\begin{equation}\label{fp} \lim_{t\rightarrow -\infty} g(t) = g^* =
\frac{\epsilon}{3}+\mathcal{O}(\epsilon^2)\end{equation} which in
turn implies that \begin{eqnarray} \lim_{t \rightarrow -\infty}
\gamma(t) & = & \eta =
\frac{\epsilon^2}{54}+\mathcal{O}(\epsilon^3)\label{anodim} \\
\lim_{t\rightarrow -\infty} v(t) & = & v(0)~ e^{(z-1) t}
\label{dynanodim}\end{eqnarray} \noindent where we introduced the
new \emph{dynamical} critical exponent \begin{eqnarray} z & = &
1+\frac{1}{2}(\eta_t-\eta) +\mathcal{O}(\epsilon^2)\label{zdyn}
\cr \cr \eta_t & = & \frac{2g^{*2}}{3\epsilon}=\frac{2
\epsilon}{27} +\mathcal{O}(\epsilon^2) \label{etatemporal}
\end{eqnarray} Therefore, in the asymptotic infrared limit we find that
\begin{equation}\label{asIRlim} \Gamma^{(2)}\left(e^t p,\frac{e^t
s}{v},g,\mu\right)= e^{t\left(2-\eta \right)}~\Gamma^{(2)}\left(
p,\frac{s~ e^{(1-z) t}}{v(0)},g^*,\mu\right)\end{equation} It is
convenient to re-define $p e^t= P~;~s e^t=S$ to find
\begin{equation}\label{asIRlimred}
\Gamma^{(2)}\left(P,\frac{S}{v},g,\mu\right)= e^{t\left(2-\eta
\right)}~\Gamma^{(2)}\left( Pe^{-t},\frac{S}{v(0)} e^{-t}~
 e^{(1-z) t},g^*,\mu\right)\end{equation}
 and finally writing $P=\mu e^t$ and using the property
 (\ref{dimlessgamma}) we find the scaling form in the infrared
 limit
 \begin{equation}\label{scalefunc}\Gamma^{(2)}\left(P,\frac{S}{v},g,\mu\right)= \mu^2
 \left[\frac{P}{\mu}\right]^{2-\eta} \Phi\left(\vartheta \right)\quad ; \quad
 \vartheta= \left(\frac{S}{v(\mu)\mu^{1-z}P^{z}}\right)^2 \end{equation}
The solution of the RG equation clearly shows that the two point
function in the infrared limit is a \emph{scaling} function of the
ratio $s/v(\mu)\mu^{1-z}p^{z}$ highlighting the role of the
\emph{new dynamical exponents} $z$ given  by eqn. (\ref{zdyn})
with $\eta_t$ to lowest order given by eqn. (\ref{etatemporal}).

The emergence of the new dynamical exponent $z$ is a consequence
of the anisotropic renormalization between momentum and frequency,
or space and time manifest in the renormalization of the speed of
light. This novel phenomenon can be traced back to the different
role played by time (compactified) and space in the Euclidean
formulation at finite temperature. A similar anisotropic rescaling
emerges in a different context: a Heisenberg ferromagnet with
correlated impurities\cite{boyaniso} with similar renormalization
group results.

While the formal solution does not yield the  function $\Phi$, we
can find it by matching to the lowest order perturbative expansion
(\ref{2ptren}) when the coupling is at the non-trivial fixed
point. From the form of the of the perturbative renormalized two
point function given by eqn. (\ref{2ptren}) and assuming the
exponentiation of the leading logarithms via the renormalization
group near the non-trivial fixed point
\begin{equation}\label{2ptresum} \Gamma^{(2)}_R(p,s;g^*)=
p^2\left[1-\eta\ln\left( \frac{p}{\mu}\right)
\right]+\frac{s^2}{v_R^2}\left[1- \eta_t \ln\left(
\frac{s}{v_R\mu}\right)\right] \thickapprox p^{2-\eta} \mu^{\eta}
+ \left(\frac{s}{v_R}\right)^{2-\eta_t}\mu^{\eta_t} \end{equation}
\noindent which can be immediately written in the scaling form
\begin{eqnarray}\label{scalingform} \Gamma^{(2)}_R(p,s;g^*) & \sim &
p^{2-\eta}\mu^{\eta}\left[1+
\left(\frac{s}{v(\mu)\mu^{1-z} p^{z}}\right)^{2-\eta_t} \right] \nonumber \\
z & = & \frac{2-\eta}{2-\eta_t} \simeq 1+\frac{1}{2}(\eta_t-\eta)
\end{eqnarray} Clearly this form coincides with the scaling solution of
the renormalization group and the perturbative expansion in the
regime in which it is valid. We note, however, that in the
computation of $\eta_t$ we have neglected contributions to the
renormalization of the velocity $v$ of $\mathcal{O}(g^3/\epsilon)$
which would appear at next order and would lead to an
$\mathcal{O}(\epsilon^2)$ contribution to $\eta_t$, which is of
the same order as $\eta$ in $z$. Thus consistently we must neglect
the contribution of $\eta$ to the dynamical exponent $z$ which to
lowest order is therefore \begin{equation} \label{dynexpz}
z=1+\frac{\eta_t}{2}+\mathcal{O}(\epsilon^2)=
1+\frac{\epsilon}{27} +\mathcal{O}(\epsilon^2) \; . \end{equation}
\subsubsection{Quasiparticles and critical slowing down:}

The quasiparticle structure of the theory is obtained from the
Green's function
$G^{-1}(p,\omega)=\Gamma^{(2)}(p,s=-i\omega+0^+,g^*)$. In
particular the dispersion relation and the width of the
quasiparticle are obtained from the real and imaginary parts,
respectively.

While the general solution of the RG equation does not determine
the scaling function $\Phi$ in eqn. (\ref{scalefunc}) the fact
that it is only a function of the scaling ratio $\vartheta$,
allows to extract the quasiparticle structure. The analytic
continuation $s\rightarrow -i\omega+0^+$ leads to the analytic
continuation of the scaling variable $\vartheta \rightarrow
-\varpi^2-i~\mathrm{sign}(\varpi)~0^+$ with \begin{equation}
\label{analcont} \varpi= \frac{\omega}{v(\mu)\mu^{1-z}p^{z}}
\end{equation} Writing the scaling function $\Phi$ analytically
continued in terms of the real and imaginary parts
$\Phi(\vartheta=-\varpi^2-i~\mathrm{sign}(\varpi)~0^+)=
\Phi_R(\varpi)+i~\Phi_I(\varpi)$, the position of the
quasiparticle pole corresponds to the value of $\varpi$ for which
the real part vanishes. Call this dimensionless real number
$\varpi^*$, hence it is clear that the dispersion relation for the
quasiparticles obeys \begin{equation} \label{disprel} \omega_p =
\varpi^* \; v(\mu) \; \mu^{1-z} \; p^{z} \end{equation}
Furthermore, \emph{assuming} that $\Phi_R$ vanishes linearly at
$\varpi^*$  we can write the Green's function near the position of
the pole in the form \begin{equation} \label{GF}
 G(\omega,p) \simeq  \frac{1}{  \mu^2
 \left[\frac{p}{\mu}\right]^{2-\eta} }~
\frac{1}{ (\varpi-\varpi^*)\Phi^{\prime}_R(\varpi^*)+ i \Phi_I
(\varpi^*) } \end{equation} Alternatively, we can write the RG
improved propagator, near the quasiparticle pole in the
Breit-Wigner form
\begin{equation} \label{BW} G_{BW}(\omega,p) \sim
\frac{\mathcal{Z}_p}{\omega-\omega_p +i~\Gamma_p} \end{equation}
\noindent with the dispersion relation, residue at the
quasiparticle pole and quasiparticle width given by
\begin{eqnarray} \omega_p & = & \varpi^* \; v(\mu)\; \mu^{1-z}\;
p^{z}
\label{quuasipole} \\
v_g & = & (z)\varpi^* \; v(\mu)\; \mu^{1-z}\; p^{z-1}
\label{quasivg}
\\
 \mathcal{Z}_p & = &
\frac{v(\mu) \; \mu^{1-z} \; p^{z}}{\mu^2 \;
\left[\frac{p}{\mu}\right]^{2-\eta}
\Phi^{\prime}_R(\varpi^*)}\label{Zpole} \\
\Gamma_p & = & \frac{\Phi_I (\varpi^*) \; v(\mu) \;\mu^{1-z}
\;p^{z}}{\Phi^{\prime}_R(\varpi^*)} \equiv  \frac{\omega_p \Phi_I
(\varpi^*)}{\varpi^* \Phi^{\prime}_R(\varpi^*)} \label{widthRG}
\end{eqnarray} The imaginary part $\Phi_I(\varpi)$ must be proportional
to the anomalous dimensions, hence perturbatively small in the
$\epsilon$ expansion (this will be seen explicitly below to lowest
order).

The definite values for $\varpi^*$,
$\Phi_R(\varpi^*);\Phi_I(\varpi^*)$ must be found by an explicit
calculation. However, the above quasiparticle properties, such as
the position of the pole, group velocity, residue and width are
\emph{universal} in the sense that they only depend on the fixed
point theory. For a positive dynamical exponent $z$ the above
analysis reveals a vanishing group velocity and width for
long-wavelength quasiparticles at the critical point.

Furthermore the expression for the width given by (\ref{widthRG})
not only displays the phenomenon of critical slowing down, i.e,
the width of the quasiparticle vanishes in the long-wavelength
limit, but also the validity of the quasiparticle picture, since
$\Gamma_p/\omega_p \ll 1$ in the $\epsilon$ expansion.

\vspace{5mm}

{\bf Threshold singularities:} While we have assumed above that
the real part of the scaling function vanishes linearly at the
quasiparticle pole, this need not be the general situation. It is
possible that the real part vanishes with an anomalous power law,
i.e, \begin{equation} \label{anomavani} \Phi_R(\Omega) \sim
|\Omega-\Omega^*|^{1+\chi} \quad ; \quad \chi =
\chi^{(1)}\epsilon+\chi^{(2)}\epsilon^2 + \cdots \end{equation} In
this case a quasiparticle width cannot be defined as the residue
will either vanish or diverge depending on the sign of $\chi$. It
is also possible that $\Phi_I(\Omega)$ also vanishes with an
anomalous power at $\Omega^*$. We refer to these cases as
threshold singularities and we will find below an example of this
case. Another example of this situation has been found in dense
QCD as a result of the breakdown of the Fermi liquid theory in the
normal phase\cite{nflqcd}. Clearly only a detailed calculation of
the scaling functions can reveal whether it vanishes linearly or
with an anomalous power law at $\Omega^*$. The set of
quasiparticle properties given above
(\ref{quuasipole}-\ref{widthRG}) are only valid provided the real
part vanishes \emph{linearly}.

\vspace{5mm}

We can go further and find the explicit form of the scaling
function by focusing on the renormalization group improved
propagator obtained in lowest order in the $\epsilon$ expansion
given by eqn. (\ref{scalingform}).

The analytic continuation to real frequencies of the RG improved
two point function to lowest order given by (\ref{scalingform})
leads to \begin{equation}\label{greensfuncRG} G^{-1}(p,\omega) =
p^{2-\eta}\mu^{\eta}\left[1-\left(\frac{|\omega|}{v(\mu)\mu^{1-z}p^{z}}\right)^{2-\eta_t}
\left(1+i  \frac{\pi \eta_t}{2} \; \mathrm{sign}(\omega)\right)
\right] \end{equation} \noindent where we have approximated
$\cos(\frac{\pi \eta_t}{2})\approx 1~;~ \sin(\frac{\pi \eta_t}{2})
\approx \frac{\pi \eta_t}{2}$ to lowest order.

From this expression we see that the dispersion relation
$\omega_p$,
 group velocity $v_g(p)$
 width $\Gamma(p)$    quasiparticles residue $\mathcal{Z}_p$ are of
 the form given
 by equations (\ref{quuasipole}-\ref{widthRG}) with $\varpi^*=1$, and
 with the following explicit
 expressions to leading order in the $\epsilon$ expansion
\begin{eqnarray}
|\omega_p| & = &   v(\mu)\,  \mu^{1-z}\,  p^{z} \\
 v_g(p) & = & z~ v(\mu)\,  \mu^{1-z}\,  p^{z-1} \label{groupvelo} \\
 \Gamma(p) & = & \frac{\pi \eta_t}{4}\,  v(\mu)\,  \mu^{1-z}\,  p^{z}
 \equiv \frac{\pi \eta_t}{4}\, |\omega_p|
 \label{width} \\
 \mathcal{Z}_p & = & \frac12 \, v(\mu) \, \mu^{1-z}
 p^{z-1} \label{residue}
 \end{eqnarray}

\noindent with $z$ given by (\ref{dynexpz}) above.

 Several important features of these expressions must be
 highlighted

 \begin{itemize}
\item{The dispersion relation (\ref{disprel}) features an
\emph{anomalous dimension} given by the dynamical exponent
$z\approx 1+\eta_t/2 = 1+\epsilon/27$. The product
$v(\mu)\mu^{1-z}$ is a renormalization group invariant as can be
seen from equations (\ref{betav}) and (\ref{betav2lup}) evaluated
at the fixed point. Thus, all of the above quantities that
describe the physical quasiparticle properties are manifestly
renormalization group invariant.}

\item{ The group velocity (\ref{groupvelo}) \emph{vanishes} in the long wavelength limit as a power law
completely determined by the dynamical anomalous dimension $z$.
This feature highlights the collective aspects of the
long-wavelength  excitations.  }

\item{{\bf Critical slowing down:} is explicitly manifest in the width $\Gamma(p)$ since
$\Gamma(p)\rightarrow 0$ as $p\rightarrow 0$. Furthermore we also
emphasize the validity of the quasiparticle picture, the ratio
$\Gamma(p)/\omega_p \approx \pi \eta_t/2 \sim
\mathcal{O}(\epsilon) \ll 1$. Thus, the quasiparticles are narrow
in the sense that their width is much smaller than the position of
the pole.  Even considering $\epsilon =1$ corresponding to
dynamical critical phenomena in three spatial dimensions
$\Gamma_p/\omega_p \sim 0.1$ }

 \end{itemize}

Thus, we see that the renormalization group resummation has led to
a consistent quasiparticle picture, but in terms of a dispersion
relation that features an anomalous dimension and a group velocity
that vanishes in the long-wavelength limit. Obviously these
features of the quasiparticles cannot be extracted from a naive
perturbative expansion.

\subsection{Away from the critical point:~$T\gtrsim
T_c$}\label{sec:awayTc}

Having studied the quasiparticle aspects at the critical point, we
now turn our attention to their study slightly away from the
critical point. The critical region of interest is $|T-T_c| \ll
T_c$.  Critical behavior in the broken symmetry phase near the
critical point with $T\lesssim T_c$ will be studied elsewhere with
particular attention on the critical dynamics of Goldstone bosons.
In this article we restrict our attention to the normal phase near
the critical point.

In the Lagrangian density (\ref{lagrangianeuclidean}) the term
$M^2(T)$ is the \emph{exact} mass (rather the exact inverse
susceptibility)  determined by the condition (\ref{massofT}) and
the counterterm $\delta m^2(T)$ is adjusted consistently in
perturbation theory to fulfil this condition. However, to relate
the mass to the departure away from the critical temperature it is
more convenient to re-arrange the perturbative expansion in a
manner that displays explicitly the departure from $T_c$. This is
achieved as follows. Consider the one loop contribution to the
self-energy in the \emph{massless} theory in $d$ -spatial
dimensions: \begin{equation} \label{masslup} \Sigma^{(1)} =
-\frac{\lambda T}{2} \sum_{m} \int \ddbarq \frac{1}{q^2 +
\omega^2_m} = -\lambda T^{d-1} \mathcal{A}(d) +
\mathrm{z.T.t}\end{equation} \noindent with $\mathcal{A}(d)$ only
depends on the spatial dimensionality and $\mathrm{z.T.t}$ stands
for zero temperature terms. It is convenient  to group this
contribution with the bare mass term in the Lagrangian in the form
\begin{equation} \label{masT} m^2(T)= -m^2_0-\Sigma^{(1)}= \lambda
T^{d-1} \mathcal{A}(d)-  m^2_R(0) \approx  a (T-T_c)~~
\mathrm{for}~ T \sim T_c = \left[ \frac{m^2_R}{\lambda
\mathcal{A}(d)}\right]^\frac{1}{d-1} \end{equation} \noindent
where the zero temperature contributions (denoted by
$\mathrm{z.T.t}$ in eqn. (\ref{masslup})) had been absorbed in
$m^2_R(0)$. We re-organize the perturbative expansion by
re-writing the Euclidean Lagrangian in the form \begin{equation}
\mathcal{L}_E= \frac{1}{2} \frac{(\partial_{\tau} \Phi)^2}{v^2_0}+
\frac{1}{2} (\triangledown \Phi)^2+m^2(T)\Phi^2 +
\frac{\lambda_R}{4!}\Phi^4 +\frac{1}{2} \delta m^2(T) \Phi^2+
\frac{\delta \lambda}{4!} \Phi^4
\label{lagrangianeuclidean2}\end{equation} \noindent where now the
counterterm $\delta m^2(T)$ is simply the one loop tadpole diagram
evaluated for zero mass given by (\ref{masslup}), it is of
$\mathcal{O}(\lambda)$ and is included in the perturbative
expansion consistently.

To one-loop order the two point function is now given by

\begin{eqnarray} \label{gamma21lup} && \Gamma^{(2)}(p,s) = p^2 +
\frac{s^2}{v^2_0}+m^2(T)-\Sigma_S \nonumber \\
&& \Sigma_S = (\Sigma-\delta m^2(T)) = \frac{\lambda T}{2}\sum_m
\int \ddbarq \frac{m^2(T)}{\left(q^2+\omega^2_m \right)\left(
q^2+\omega^2_m+m^2(T)\right)} \end{eqnarray} The integral above is
of the typical form as those studied in the previous sections (see
(\ref{epsiHiT})). Separating the $m=0$ Matsubara contribution from
the $m\neq 0$ for which we can set $m^2(T) \propto (T-T_c) \approx
0$ for $T-T_c \ll T_c$, we obtain

\begin{equation}\label{intg}
\sum_m \int \ddbarq \frac{m^2(T)}{\left(q^2+\omega^2_m
\right)\left[ q^2+\omega^2_m+m^2(T)\right]} = m^2(T)\left[
\frac{\Gamma(\frac{\epsilon}{2})\Gamma(1-
\frac{\epsilon}{2})}{(4\pi)^{\frac{d}{2}}\Gamma(\frac{d}{2})}
m^{-\epsilon}(T) + \mathcal{C}(d) \;  T^{-\epsilon}\right]
\end{equation}

Again, for $\epsilon >0$ we can neglect the second term in the
brackets in the limit $T \gg m(T)$. Expanding in $\epsilon$ to
obtain the lowest order contribution consistently in the
$\epsilon$ expansion, we obtain the two point function to one loop
order \begin{equation}\label{gama2mT} \Gamma^{(2)}(p,s) \approx
p^2 +
\frac{s^2}{v^2_0}+m^2(T)\left[1-\frac{g(\mu)}{\epsilon}+\frac{g(\mu)}{2}\ln\left(\frac{m^2(T)}{\mu^2}\right)
\right] \end{equation} The renormalized mass parameter $m_R(T)$ is
defined by \begin{equation} \label{renmas} Z_{\phi}~m^2(T)= Z_m~
m^2_R(T) \end{equation} and $Z_m$ is fixed by the renormalization
condition
\begin{equation} \Gamma^{(2)}(p=0,s=0,m^2_R=\mu^2)=\mu^2 \end{equation} Since
$Z_{\phi}$ receives corrections at $\mathcal{O}(g^2)$ we choose
$Z_m$ to lowest order in the $\epsilon$ expansion to be

\begin{equation} \label{zmass} Z_m= 1+\frac{g(\mu)}{\epsilon} +
\mathcal{O}(g^2,g\epsilon) \end{equation}

\subsubsection{ Static aspects:}

Before we embark on a full discussion of the dynamics away from
the critical point, it proves convenient and illuminating to
discuss the static aspects first. In particular since we will
study the $p=0$ case but $T\neq T_c$ a relevant quantity is the
inverse susceptibility $\chi^{-1}(T)$, which  is defined as
  \begin{equation} \label{corrlength}
  \chi^{-1}(T)=M^{2}(T) = \Gamma^2_R(p=0;s=0) \end{equation}
  which near the critical point and the non-trivial fixed point
  $g^*$ given by eqn. (\ref{fp}) is given by
  \begin{equation}\label{corrresum} M^{2}(T\sim T_c) \approx
  m^2_R(T)\left[1+ \frac{g^*}{2}\ln\left(\frac{m^2_R(T)}{\mu^2}\right)
  \right]\approx \left[ m^2_R(T) \right]^{1+\frac{g^*}{2}} \end{equation}
\noindent where we anticipated an exponentiation of the leading
logarithms via the renormalization group, which will be borne out
by the renormalization group analysis  below. Recalling that
$m^2(T) \propto |T-T_c|$ by eqn. (\ref{masT}), we find
\begin{equation} \label{corrfin} \chi^{-1}(T\sim T_c) \propto
|T-T_c|^{\gamma} ~;~ \gamma = 1+ \frac{\epsilon}{6}+\cdots
\end{equation} The critical exponent $\gamma$ is seen to be the
correct one \cite{ma,amit,cardy,zj}.

Just as in the case of the theory at the critical point studied
above, we now study the dynamics of the theory in an $\epsilon$
expansion and implement a resummation of the leading infrared
divergences via the renormalization group.

\subsubsection{Dynamics away from the critical point}

As argued above the leading infrared behavior is obtained by
setting the internal Matsubara frequencies to zero in the two
loops self energy.  In $d=4-\epsilon$ spatial dimensions, the
self-energy at two loops is \begin{equation} \Sigma^{(2)}( p ;s;
m(T))= \frac{\lambda^2 T^2}{6}\int \ddbarq \int \ddbark
\frac{1}{[q^2+m^2(T)][k^2+m^2(T)][(q+k+p)^2+\frac{s^2}{v^2_0}+m^2(T)]}
\label{sig2lTg} \end{equation} The loop integrals are  evaluated
by introducing two Feynman parameters leading to \begin{eqnarray}
\label{ints} \Sigma^{(2)}(p ;s; m(T)) = g^2(\mu)
\Gamma(-1+\epsilon)\mu^{2\epsilon} \int^1_0 dx \int^1_0 dy &&
[x(1-x)]^{\frac{\epsilon}{2}-1}y^{\frac{\epsilon}{2}-1}
\left[x(1-x)y(1-y)p^2+\right. \nonumber \\ && \left.
+(1-y)x(1-x)m^2(T)+y m^2(T)+yx\frac{s^2}{v^2_0}
\right]^{1-\epsilon} \end{eqnarray} It is convenient to separate
the static contribution from the dynamical part by writing
\begin{equation} \Sigma^{(2)}(p ;s; m(T))= \Sigma^{(2)}(p ;0;
m(T))+ \tilde{\Sigma}^{(2)}(p ;s; m(T))~~;~~\tilde{\Sigma}^{(2)}(p
;s; m(T))\equiv {\Sigma}^{(2)}(p ;s; m(T))-{\Sigma}^{(2)}(p ;0;
m(T)) \end{equation} The static contribution ${\Sigma}^{(2)}(p ;0;
m(T))$ leads to wave function renormalization a renormalization of
the mass and to
 $\mathcal{O}(\epsilon^2)$ corrections to the static anomalous
 dimensions, which will be neglected to leading order in the
 $\epsilon$ expansion. The second,
dynamical contribution is obtained consistently in an $\epsilon$
expansion: the regions of the integrals in the Feynman parameters
that lead to inverse powers of $\epsilon$ in an $\epsilon$
expansion are $x\sim 0,1$ and $y \sim 0$. The contributions of
these regions can be isolated  by partial integration, and after
some straightforward algebra we find \begin{equation}
\label{tilsig} \tilde{\Sigma}^{(2)}(p ;s; m(T))=
\frac{g^2(\mu)}{6\epsilon}\left[-\frac{2}{\epsilon}\frac{s^2}{v^2_0}+2\frac{s^2}{v^2_0}
\ln\left(\frac{m^2(T)}{\mu^2}\right)+
 2\left(m^2(T)+\frac{s^2}{v^2_0}\right)\ln\left(1+
\frac{s^2}{v^2_0 m^2(T)} \right) \right]
+\mathcal{O}(\epsilon^0,\epsilon) \end{equation} Thus, putting
together the one loop contribution found previously and the two
loop contribution found above, the two point function at zero
spatial momentum but away from the critical point is found to be
\begin{eqnarray} \label{twopointawayTc} \Gamma^{(2)}(p;s;m(T)) & = & p^2
\left[1+ \frac{g^2(\mu)}{12\epsilon}+ \mathcal{O}(g^2\epsilon^0)
\right]+ \frac{s^2}{v^2_0}+m^2(T)\left[1-\frac{g(\mu)}{\epsilon}+
\frac{g(\mu)}{2}\ln\left(\frac{m^2(T)}{\mu^2}\right) \right]+ \nonumber \\
&& \frac{g^2(\mu)}{6\epsilon}\left[\frac{2}{\epsilon}
\frac{s^2}{v^2_0}-2\frac{s^2}{v^2_0}
\ln\left(\frac{m^2(T)}{\mu^2}\right)-
2\left(m^2(T)+\frac{s^2}{v^2_0}\right)\ln\left(1+ \frac{s^2}{v^2_0
m^2(T)} \right) \right] \nonumber \\ &&
+\mathcal{O}(g^2,g^2\epsilon) \end{eqnarray} \noindent where we
have neglected logarithmic corrections that will exponentiate to
anomalous dimensions of $\mathcal{O}(\epsilon^2)$ for momentum
dependent terms. We have only displayed the contribution that will
be cancelled by wave function renormalization just as in the
critical case.

Obviously the $m^2(T)=0$ limit coincides with the two point
function at the critical point (\ref{gammaeps}) to leading order
$\mathcal{O}(\epsilon)$. In the above expression we have not
included the two loop contribution to the static $s=0$ part, since
it will lead to an $\mathcal{O}(\epsilon^2)$ correction to the
critical exponent for the correlation length (inverse
susceptibility).

The renormalization conditions for the two point function away
from the critical point are now summarized as follows:

\begin{eqnarray}\label{rencondaway} && \Gamma^{(2)}_R(p,s,v_R;m_R(T))=
Z_{\phi}~\Gamma^{(2)}(p,s,v_0;m(T)) \nonumber \\
&&v^2_R=v^2_0~\frac{Z_v}{Z_{\phi}} ~~;~~  Z_{\phi}~m^2(T)= m^2_R(T)~Z_m \nonumber \\
 && \left.\frac{\partial \Gamma^{(2)}_R} {\partial
p^2}\right|_{p^2=\mu^2,\frac{s^2}{v^2_R}=\mu^2}= 1 ~~;~~
\left.\frac{\partial \Gamma^{(2)}_R}{\partial
s^2}\right|_{p^2=\mu^2,\frac{s^2}{v^2_R}=\mu^2}= \frac{1}{v^2_R}\\
&& \Gamma^{(2)}_R(p=0,s=0;m^2_R(T)=\mu^2)=\mu^2 \label{newcond}
   \end{eqnarray}
\noindent along with the renormalization conditions on the four
point function (\ref{couprenor}).  To leading order in the
$\epsilon$ expansion the renormalization constants
$Z_{\phi}\,,Z_v\,,Z_m$ are given by (\ref{zphi}),( \ref{zvel}) and
(\ref{zmass}) respectively.

Thus, we find the renormalized two point function at two loop
order and to leading order in the $\epsilon$ expansion (since $g^*
\sim \epsilon$) \begin{eqnarray} \label{rentwopointawayTc}
\Gamma^{(2)}_R(p;s;m_R(T)) & = & p^2+
m^2(T)\left[1+\frac{g(\mu)}{2}\ln\left(\frac{m^2_R(T)}{\mu^2}\right)
\right]+ \frac{s^2}{v^2_R}\left[1- \frac{g^2(\mu)}{3\epsilon}
\ln\left(\frac{m^2_R(T)}{\mu^2}\right)\right] \nonumber \\
 && -\frac{g^2(\mu)}{3\epsilon}\left(m^2_R(T)+
\frac{s^2}{v^2_0}\right)\ln\left(1+ \frac{s^2}{v^2_0 m^2_R(T)}
\right) +\mathcal{O}(g^2,g^2\epsilon)
 \end{eqnarray}
Since $m^2_R(T)$ has dimension two  it is convenient to introduce
the dimensionless quantity \begin{equation} \bar{m}^2=
\frac{m^2_R(T)}{\mu^2} \label{dimensionlessmass} \end{equation}
\noindent and the corresponding renormalization group beta
function \begin{eqnarray} \beta_{\bar m} & = & \left.\mu
\frac{\partial \bar{m}^2}{\partial \mu}\right|_{m_0,T,\lambda_0} =
(\gamma_{\bar m}
-2)~ \bar{m}^2\label{betamass} \\
\gamma_{\bar m}  & = & g + \mathcal{O}(g^2,g\epsilon)
\label{gammamass} \end{eqnarray}

\noindent where we have used (\ref{renmas}) and (\ref{zmass}).

The renormalization group equation for the $N$-point function away
from the critical point is now given by
\begin{equation}\label{rgawayTc} \left[ \mu {\partial \over
\partial\mu} + \beta_{g}\frac{\partial}{\partial g} +
\beta_v\frac{\partial}{\partial v}+\beta_{\bar
m}\frac{\partial}{\partial {\bar m}^2} - {N \over 2} \; \gamma
\right]\Gamma^{(N)}\left(p_1,\frac{s_1}{v};
p_2,\frac{s_2}{v}\ldots,p_N;\frac{s_N}{v}; g,\bar{m},\mu\right) =
0 \end{equation} The new ingredient as compared to the critical
case (\ref{egr}) is the dependence on $\bar{m}$. Following the
same steps as for the critical case, we now find that the solution
of the renormalization group equation for the two point function
obeys \begin{equation}\label{rgsolawayTc} \Gamma^{(2)}\left(e^t
p,\frac{e^t s}{v},g,\bar{m},\mu\right)= e^{\int_0^t
dt'\left(2-\gamma(t')\right)}\Gamma^{(2)}\left(
p,\frac{s}{v(t)},g(t),\bar{m}^2(t),\mu\right)\end{equation}
\noindent with $\bar{m}(t)$ the solution of the differential
equation
\begin{equation} \frac{\partial \bar{m}^2(t)}{\partial t}=
\beta_{\bar m}(g(t),v(t),\bar{m}(t)) \label{betambar}
\end{equation} \noindent with the initial condition
\begin{equation}\label{iniconmofT}
\bar{m}^2(0)=\frac{m^2_R(T,\mu)}{\mu^2} \propto
|T-T_c(\mu)|\end{equation} In the infrared the coupling is driven
to the non-trivial fixed point $g^*=\epsilon/3$ and
\begin{equation} \bar{m}^2(t) \rightarrow \bar{m}^2(0)~
e^{(\gamma^*_{\bar m}-2)t} \quad ;\gamma^*_{\bar m} =
\frac{\epsilon}{3} \end{equation} Just as in the solution of the
renormalization group equation at criticality near the fixed point
(\ref{asIRlim},\ref{asIRlimred}), introducing $pe^t\equiv
P\,;\,se^t\equiv S$ we now find
\begin{equation}\label{asIRlimredTnoTc}
\Gamma^{(2)}\left(P,\frac{S}{v},g,\mu\right)= e^{t\left(2-\eta
\right)}~\Gamma^{(2)}\left( Pe^{-t},\frac{S}{v(0)} e^{-t}~
 e^{(1-z) t},g^*,\bar{m}^2(0) e^{(\gamma^*_{\bar m}-2)t},\mu\right)\end{equation}
Following the analysis of the critical case, and the scaling
property (\ref{dimlessgamma}) and writing $P=\mu e^t$ we find the
following scaling form \begin{equation}
\label{scalform}\Gamma^{(2)}\left(P,\frac{S}{v},g,\mu\right)=
\mu^2
 \left[\frac{P}{\mu}\right]^{2-\eta}
 \Phi\left(\frac{S}{v(\mu)\mu^{1-z} P^{z}};P\xi
 \right) \quad;\quad \xi= \frac{1}{\mu}\left[\frac{ m^2_R(T,\mu)}{\mu^2}\right]^{\frac{1}{\gamma^*_{\bar m}-2}}\end{equation}
 \noindent $\xi$ is therefore identified with the
 \emph{correlation length}\cite{ma,amit,cardy,zj}
\begin{equation}\label{corre}\xi \sim |T-T_c|^{-\nu} \quad; \quad \nu=
\frac{1}{2-\gamma^*_{\bar m}} \sim
\frac{1}{2}+\frac{\epsilon}{12}+\cdots \end{equation}
  It is important to note at this stage
 that the correlation length $ \xi$
 is a renormalization group invariant, as can be easily checked by
 using (\ref{dimensionlessmass}) with the renormalization group beta function
 (\ref{betamass}).

 To study the limit of zero spatial momentum it is more convenient to
 rewrite the above scaling solution in the following form
\begin{equation} \label{scalform2}\Gamma^{(2)}\left(P,\frac{S}{v},g,\mu\right)=
\mu^2
 \left[\frac{\xi}{\mu}\right]^{-(2-\eta)}
 \Psi\left(\frac{S~\xi^{z}}{v(\mu)\mu^{1-z}};P\xi
 \right)\end{equation}
 From the definition of the inverse susceptibility
 $M^2(T)=\chi^{-1}(T)=\Gamma^{(2)}(p=0,s=0)$ we find the known
 result\cite{ma,amit,cardy,zj,dombgreen}
\begin{equation} \chi^{-1}(T) \propto |T-T_c|^{-\gamma}\quad ; \quad \gamma =
\frac{2-\eta}{2-\gamma^*_{\bar
 m}}=\nu(2-\eta) \end{equation}
Furthermore the two-point function  is a function of two
renormalization group invariant, dimensionless scaling
 variables

\begin{equation} \label{finfor} \Gamma^{(2)}(p,s,m^2_R(T,\mu))  =  \mu^2
\left[\frac{m^2_R(T,\mu)}{\mu^2}\right]^{\frac{2-\eta}{2-\gamma^*_{\bar
 m}}}\Psi\left(\varphi, \delta   \right) \end{equation}
 \noindent with
\begin{eqnarray} \varphi & = &
 \left[\frac{s}{v(\mu)\mu }\right]^2\left[\frac{m^2_R(T)}{\mu^2}\right]^{\frac{2z}{\gamma^*_{\bar m}-2}}\equiv
 \left[\frac{s\,\xi^{z}}{v(\mu)\mu^{1-z}} \right]^2\label{varphi}
 \\ \delta & = & \frac{p^2}{\mu^2}\left[\frac{ m^2_R(T,\mu)}{\mu^2}\right]^{\frac{2}{\gamma^*_{\bar m}-2}} \label{delta} \equiv
 (p\,
  \xi)^2
 \end{eqnarray}
 The renormalization condition (\ref{newcond}) determines that
 $\Psi(0,0)=1$.

 We can now follow the arguments provided in the previous
 subsection for the critical case. Under the analytic continuation
 $s^2 \rightarrow -\omega^2-i~\mathrm{sign}(\omega)~0^+$
 \begin{equation} \label{contaway}  \varphi \rightarrow -\Omega^2-i
 \mathrm{sign}(\Omega)~0^+ \quad \quad; \quad \Omega
 =\frac{\omega \,\xi^{z}}{v(\mu)\mu^{1-z} }\end{equation}
\begin{equation} \Psi(\varphi=-\Omega^2-i
 \mathrm{sign}(\Omega)~0^+,\delta )= \Psi_R(\Omega,\delta)+i~\Psi_I(\Omega,\delta) \label{psifunc}\end{equation}
 The position of the quasiparticle pole in the two point Green's
 function corresponds to the value of $\Omega=\Omega^*(\delta)$ for which
 $\Psi_R(\Omega^*(\delta),\delta)=0$. This condition determines the
 dispersion relation of the quasiparticle  and is given by
 \begin{equation}
\omega_p= \Omega^*(\delta) \; v(\mu) \; \mu^{1-z}\; \xi^{-z}
\label{poleaway} \end{equation} \noindent this expression
emphasizes that the dispersion relation depends on $p$ through the
scaling variable $\delta= (p\,\xi)^2$.

 \emph{Assuming} that near the quasiparticle pole $\Psi_R$ vanishes \emph{linearly} the Green's function  can be
 approximated by
 \begin{equation}\label{GFaway} G(p, \omega,m_R(T)) \sim \frac{(\mu \,
 \xi)^{2-\eta}}{\mu^2}
~\frac{1}{(\Omega-\Omega^*)\Psi^{\prime}_R(\Omega^*,\delta)+i~\Psi_I(\Omega^*,\delta)}
\end{equation} \noindent where $\Psi^{\prime}_R(\Omega^*,\delta)= \partial
\Psi_R(\Omega,\delta)/\partial \Omega |_{\Omega=\Omega^*}$. Near
the quasiparticle pole we can further  write  the above expression
in the Breit-Wigner form \begin{equation} \label{BWaway} G_{BW}(p,
\omega,m_R(T)) \sim
\frac{\mathcal{Z}_p}{\omega-\omega_p+i~\Gamma_p} \end{equation}
\noindent with \begin{eqnarray} \omega_p & = & \Omega^*(\delta)~
\frac{v(\mu)\mu^{1-z}}{\xi^{z}}
\label{omegapoleTc} \\
\mathcal{Z}_p & = & \Psi^{\prime}_R(\Omega^*)\frac{v(\mu)}{\mu}
\left(\mu \, \xi\right)^{2-z-\eta} \label{residueaway} \\
 \Gamma_p &= & \frac{\Psi_I(\Omega^*)}{\Psi^{\prime}_R(\Omega^*)}\frac{v(\mu)\mu^{1-z}}{\xi^{z}}  \equiv
 \frac{\omega_p~\Psi_I(\Omega^*)}{\Omega^*(\delta)~\Psi^{\prime}_R(\Omega^*)}
  \propto |T-T_c|^{z \nu }\label{widthaway} \end{eqnarray}
\noindent where we have suppressed the dependence on the scaling
variable $\delta$ in the arguments of the real and imaginary parts
to avoid cluttering of notation. Furthermore,  we have made
explicit the combination of \emph{static and dynamic} critical
exponents using the expression given in eqn. (\ref{corre}) for the
static critical exponent $\nu$ and the dependence on the momentum
is implicit through the dependence on the scaling variable
$\delta$ of $\Omega^*$ as well as the explicit dependence of the
real and imaginary parts.

Again, the imaginary part must be proportional to the anomalous
dimensions, hence perturbatively small in the $\epsilon$
expansion. Therefore the expression for the width
(\ref{widthaway}) reveals both critical slowing down, since
$\Gamma_p \sim |T-T_c|^{z \nu }$ vanishing at $T=T_c$, \emph{and}
the validity of the
 quasiparticle picture since $\Gamma_p/\omega_p \ll 1$ in the
 $\epsilon$ expansion.

 At this point we recognize a fundamental difference with the
 Wilsonian results of reference\cite{pietroni}. While in ref.
 \cite{pietroni} the width was found to be proportional to
 $|T-T_c|^\nu$ up to logarithms, we see from (\ref{widthaway})
 that the quasiparticle width actually involves the new dynamical
 anomalous exponent $z$. The difference can be traced to the fact
 that the Wilsonian approach advocated in ref.\cite{pietroni}
  does {\em not} include two loop diagrams which are necessary to
  reveal the anisotropic renormalization through the
  renormalization of the speed of light and are directly responsible for the new dynamical anomalous exponent $z$.

 We emphasize that the above Breit-Wigner form as well as the
 quasiparticle properties rely on the assumption that the real
 part of the scaling function vanishes linearly near the
 quasiparticle pole. As emphasized before in the critical case
 this need not be the general situation, and anomalous power laws
 can lead to threshold singularities as discussed above.

 While the solution of the renormalization group leads to a
scaling form of the two point correlation function, it does {\em
not} explicitly specify the scaling function $\Psi$. However, we
can obtain the function $\Psi$ by matching the leading logarithms
to those of the perturbative expression (\ref{rentwopointawayTc})
 evaluated at the fixed
point $g^*=\epsilon/3$ to lowest order in the $\epsilon$
expansion. Matching the leading logarithms and assuming their
exponentiation  via the renormalization group   it is
straightforward to see that the two point function is given by
\begin{equation} \label{resumed2pt}
\Gamma^{(2)}(p,s,m^2_R(T,\mu))\sim \mu^2
\left[\frac{m^2_R(T,\mu)}{\mu^2}\right]^{\frac{2-\eta}{2-\gamma^*_{\bar
 m}}}\left\{\delta+\left[1+\varphi \right]^{2-z}\right\} \end{equation}
\noindent where we have used the lowest order results in the
$\epsilon$ expansion \begin{equation} \label{loworderexps}
\gamma^*_{\bar
 m} = \frac{\epsilon}{3} \quad ; \quad z= 1+\frac{\epsilon}{27}
 \quad ;  \quad \eta= \mathcal{O}(\epsilon^2) \end{equation}
 \noindent and kept consistently the lowest
 $\mathcal{O}(\epsilon)$ in the exponentiation of the leading
 logarithms leading to (\ref{resumed2pt}). Thus, we obtain the
 lowest order result for the scaling function
 \begin{equation} \label{scalefuncaway}
 \Psi(\varphi,\delta)=\delta+\left[1+\varphi
 \right]^{2-z}\end{equation}
We can now obtain an explicit form of the real and imaginary parts
of the scaling function that enter in the quasiparticle
parameters. This is achieved by performing the analytic
continuation (\ref{contaway}) which leads to \begin{eqnarray}
\Psi_R(\Omega) + i \Psi_I(\Omega) & = & \delta+ \left[1-\Omega^2
-i~\mathrm{sign}(\Omega)~0^+\right]^{2-z} \nonumber \\
& = & \delta-|\Omega^2-1|^{2-z}\left[1+i~\pi~(z-1)~
\mathrm{sign}(\Omega)\Theta(\Omega^2-1)\right] \label{anaconti}
 \end{eqnarray}
For $p=0$ i.e, $\delta=0$ we see that both the real and imaginary
part of the scaling function vanish at $\Omega^*=1$ \emph{with an
anomalous power law} providing an explicit example of the case of
threshold singularities mentioned above.

For $p\neq 0$ and $T\neq T_c$ we find a quasiparticle pole at
\begin{equation} \label{quasiawayTc} \Omega^{*2}= 1+
\delta^{\frac{1}{2-z}}\sim 1+(p\, \xi)^{2z} \end{equation}
\noindent where we have approximated the anomalous dimension by
its leading order in $\epsilon$ using $z=1+\epsilon/27$. From this
expression for $\Omega^*$ we obtain the dispersion relation for
quasiparticles
\begin{equation}\label{disperaway} \omega^2_p =
v^2(\mu)\mu^2\left\{\left[\frac{m^2_R(T,\mu)}{\mu^2}\right]^{2z\nu}+\left[\frac{p^2}{\mu^2}
\right]^{z} \right\}\end{equation} \noindent with $\nu$ given by
eqn. (\ref{corre}), in particular we find that the frequency of
zero momentum quasiparticles $\omega_{p=0}\propto ~ |T-T_c|^{z\nu
}$. Obviously at $T-T_c$ ($m_R(T)=0$) the dispersion relation
coincides with that of the critical case given by eqn.
(\ref{disprel}).  For $p\neq 0$, i.e, $\delta \neq 0$ the real
part of the scaling function vanishes linearly and the
Breit-Wigner approximation (\ref{BWaway}) near the quasiparticle
pole is valid and the relations
(\ref{omegapoleTc}-\ref{widthaway}) describe the properties of the
quasiparticles. To lowest order in the $\epsilon$ expansion we
find, using (\ref{dynexpz}) that \begin{equation}
\label{ratioaway}
 \frac{\Psi_I(\Omega^*)}{\Omega^*(\delta)\Psi^{\prime}_R(\Omega^*)} =  \frac{\pi \eta_t}{4}
 \frac{(p\,\xi)^{2z}}{1+(p\,\xi)^{2z}}\end{equation}
 \noindent for $p=0$, i.e, $\delta=0$ this ratio vanishes and
 $\Psi^{\prime}_R(\Omega^*)\propto
 |\Omega^*-1|^{1-z}_{\Omega^*=1}$ diverges
 displaying the phenomenon of threshold singularity with a divergent residue $\mathcal{Z}_p$.

For $p\neq 0$ but $T\rightarrow T_c$ ($\delta \rightarrow \infty$
) this ratio equals that of the critical case (see eqn.
(\ref{width})).

 For $T\neq T_c; p\neq 0$ we finally find the width of the long
 wavelength quasiparticles to be given to lowest order in the $\epsilon $ expansion by
 \begin{equation} \label{widthlast}
\Gamma_p \sim \frac{\pi \eta_t}{4}
 \frac{(p\,\xi)^{2z}}{1+(p\,\xi)^{2z}} \frac{v(\mu)\mu^{1-z}}{\xi^{z}}
 \left[1+\left(p \,\xi\right)^{2z}
\right]^{\frac{1}{2}} \end{equation} \noindent with the following
behavior to lowest order in the $\epsilon$ expansion
\begin{equation} \label{limits} \Gamma_p \sim \frac{\pi
\eta_t}{4}\, v(\mu) \; \mu^{-z} \left\{
\begin{array}{cc}
p^{z} &  \mathrm{for} \quad p \quad \mathrm{fixed},  T\rightarrow T_c \\
p^{2z}\xi^{z} & \mathrm{for} \quad \xi \quad \mathrm{fixed},
p\rightarrow 0
\end{array} \right.
\end{equation} Thus, critical slowing down emerges in both limits,
furthermore the validity of the quasiparticle picture is warranted
in the $\epsilon$ expansion, since $\eta_t \simeq 2(z-1) =
2\epsilon/27+\mathcal{O}(\epsilon^2) \ll 1$. \vspace{5mm}
\section{\label{sec:summary} Summary of results:}

Critical phenomena, both static and dynamic in quantum field
theory at finite temperature results in dimensional reduction
since momenta and frequencies are $p,\omega \ll T$ and the
correlation length is $\xi \gg 1/T$. The infrared physics is
dominated by the contribution of the zero Matsubara frequency in
internal loops, which in turn results in an effective coupling
$\lambda T$ in the perturbative expansion. Naive perturbation
theory at high temperature breaks down in four space-time
dimensions because of the strong infrared behavior of loop
diagrams near the critical point for long-wavelength phenomena.

We propose an implementation of the renormalization group to study
\emph{dynamical} critical phenomena which hinges upon two main
ingredients:

\begin{itemize}
\item{The leading infrared behavior near the critical point is determined
by keeping only the zero Matsubara internal frequency in the
loops. To control the infrared consistently we implement an
expansion in $\epsilon$ in $5-\epsilon$ space-time dimensions.
Dimensional reduction for long-wavelength phenomena near the
critical point results in that the perturbative expansion is in
terms of $g(\mu) \propto \lambda T \mu^{-\epsilon}$ where $\mu$ is
the scale of external momenta and frequencies in the diagram. The
renormalized effective coupling is driven to a fixed point in the
infrared which is of $\mathcal{O}(\epsilon)$. Therefore
long-wavelength phenomena can be studied  in perturbation theory
around this fixed point for $\epsilon \ll 1$. The perturbative
expansion is improved by implementing a renormalization group
resummation which reveals dynamical scaling phenomena with
anomalous dimensions. Eventually the limit of physical interest
$\epsilon \rightarrow 1$ must be studied by further Borel and/or
Pad\'e resummations.   }

\item{ The second important ingredient is the  \emph{anisotropic} scaling between space and time.
While space is infinite, at finite temperature in the Euclidean
formulation the time direction is compactified to the interval
$[0,1/T]$. We introduce a new parameter, the effective speed of
light in the medium, which is renormalized and runs with the
renormalization transformations. The infrared renormalization of
the speed of light results in a new \emph{dynamical} anomalous
exponent which determines the dispersion relation and all the
quasiparticle properties. The $\epsilon$ expansion combined with
the renormalization group leads to a consistent quasiparticle
description of long-wavelength excitations near the critical
point. }

\end{itemize}

The critical exponents, both static and dynamic are summarized
below for the critical case $T=T_c$ as well as for $T \neq T_c$
but in the symmetric phase with $T\rightarrow T_c^+$.
\begin{equation} \label{table1}
\begin{array}{|c|c|}\hline
\multicolumn{2}{|c|}{\bf Table ~1:~ Quasiparticles~ at ~T = T_c } \\
\hline \hline
  \omega_p ~\propto &  p^{z}  \\ \hline
 v_g ~ \propto &  p^{z-1} \\ \hline
 \mathcal{Z}_p ~ \propto & \ p^{z+\eta-2} \\ \hline
 \Gamma_p ~ \propto  & \eta_t ~ p^{z} \\ \hline
 \eta ~=&
\frac{\epsilon^2}{54}+\mathcal{O}(\epsilon^3)~(\mathrm{static}) \\
\hline \eta_t ~ =&  \frac{2\epsilon}{27}+\mathcal{O}(\epsilon^2)
~(\mathrm{dynamic}) \\
\hline
 z ~ =  &  1+\frac{1}{2}(\eta_t-\eta)\sim 1+\frac{\epsilon}{27}+\mathcal{O}(\epsilon^2) ~(\mathrm{dynamic}) \\ \hline
\end{array}
 \end{equation}
\begin{equation} \label{table2}
\begin{array}{|c|c|}\hline
\multicolumn{2}{|c|}{\bf Table~ 2: Quasiparticles~ at ~T \gtrsim T_c } \\
\hline \hline
  \omega^2_p ~\propto & \left[\frac{m^2_R(T,\mu)}{\mu^2}\right]^{2z\nu
}+\left[\frac{p^2}{\mu^2} \right]^{z}  \\ \hline
   \Gamma_p ~ \propto  &  \eta_t~\omega_p~
 \frac{(p\,\xi)^{2z}}{1+(p\,\xi)^{2z}} \\ \hline
 \xi \propto & |T-T_c|^{-\nu} \\ \hline
m^2_R(T) ~\propto &  |T-T_c| \\
\hline
\nu  =&  \frac{1}{2}+\frac{\epsilon}{12}+\mathcal{O}(\epsilon^2)~(\mathrm{static}) \\
\hline \eta_t ~ =&  \frac{2\epsilon}{27}+\mathcal{O}(\epsilon^2)
~(\mathrm{dynamic}) \\ \hline
 z ~ = & \simeq 1+\frac{1}{2}(\eta_t-\eta) =
1+\frac{\epsilon}{27}+\mathcal{O}(\epsilon^2) ~(\mathrm{dynamic}) \\
\hline
\end{array}
 \end{equation}
The new dynamical exponent $z$ is missed by the Wilsonian approach
advocated in reference\cite{pietroni} since two loops diagrams are
completely neglected in that approach and anisotropic rescaling of
frequency and momenta becomes manifest at two loop order and
beyond.

\vspace{2mm} {\bf Critical exponents for $O(N)$ symmetry:}

 At this stage we can generalize our results to the case of a
 scalar theory with $O(N)$ symmetry at or slightly above the
 critical point. While the static critical exponents for the
 $O(N)$ case are available in the
 literature\cite{ma,amit,cardy,zj,dombgreen}, the dynamical
 critical exponent to lowest order in $\epsilon$  can be obtained
 simply by recognizing that the symmetry factors corresponding to
 the $O(N)$ theory multiply the two loop expression for the
 self-energy by an overall factor.  From the expression
 (\ref{gammaeps}) we see that the coefficient of $s^2/v^2$ is a factor $4/\epsilon$ times
 the coefficient of $p^2$, which immediately leads to the result
\begin{equation} \eta_t = \frac{4}{\epsilon}\eta \end{equation}
 Since for the $O(N)$ theory $\eta =\epsilon^2~
(N+2)/[2(N+8)^2]+ \mathcal{O}(\epsilon^3)$ we find to lowest order
in $\epsilon$
$$
\eta_t =\epsilon ~ {2(N+2) \over (N+8)^2}+\mathcal{O}(\epsilon^2)
\; .
$$
In summary the static and dynamic critical exponents to lowest
order in the $\epsilon$ expansion for the $O(N)$ theory are given
by \begin{equation} \label{table3}
\begin{array}{|c|c|}\hline
\multicolumn{2}{|c|}{\bf Table~ 3: Critical~ exponents ~ for ~ $O(N)$. } \\
\hline \hline
\nu  =&  \frac{1}{2}+\frac{N+2}{4(N+8)}\epsilon+\mathcal{O}(\epsilon^2) ~(\mathrm{static}) \\
\hline \eta = & \epsilon^2~
\frac{N+2}{2(N+8)^2}+\mathcal{O}(\epsilon^3)~(\mathrm{static}) \\
\hline
\eta_t ~ =&  \epsilon ~\frac{2(N+2)}{(N+8)^2}+\mathcal{O}(\epsilon^2) ~(\mathrm{dynamic}) \\
\hline
 z ~ = & 1+ \epsilon ~ \frac{N+2}{(N+8)^2}+\mathcal{O}(\epsilon^2)
~(\mathrm{dynamic}) \\ \hline
\end{array}
 \end{equation}
\section{\label{sec:conclusions}Conclusions, discussion and implications}

We have studied the dynamical aspects of long-wavelength
(collective) excitations at and near the critical point in scalar
quantum field theories at high temperatures. After recognizing
that naive perturbation theory breaks down at high temperature in
the long-wavelength limit, we introduced an $\epsilon$ expansion
around $5-\epsilon$ space time dimensions combined with the
renormalization group at high temperature to resum the
perturbative series.

The effective long-wavelength theory at high temperature is
described by a non-trivial fixed point at which the correlation
functions feature scaling behavior. The anisotropy between spatial
and time coordinates in Euclidean space-time at finite temperature
leads to consider the renormalization of the speed of light,
which, in turn leads to a new \emph{dynamical} exponent $z$. All
dynamical quantities, such as the dispersion relation and widths
of long-wavelength quasiparticle (collective) excitations depend
on this new dynamical exponent, as well as the static exponents.

Our results are summarized in the tables in the previous section.

Two very important aspects emerge from this treatment: i) critical
slowing down, i.e, the relaxation rate of the quasiparticle
vanishes in the long-wavelength limit or at the critical point
with definite  anomalous dimensions determined by the new
\emph{dynamical} exponent $z$ and ii) the quasiparticle picture,
i.e, narrow widths $\Gamma_p \ll \omega_p$ is valid. The group
velocity of quasiparticles vanishes at the critical point in the
long-wavelength limit revealing the collective aspects of these
excitations. The dynamical exponent $z=1+ \epsilon ~
\frac{N+2}{(N+8)^2}+\mathcal{O}(\epsilon^2)$ describes a \emph{new
universality class} for \emph{dynamical} critical phenomena in
quantum field theory.

As mentioned in the introduction  these phenomena have
phenomenological implications for the chiral phase transition in
the Quark Gluon plasma with potential observational consequences
if long-wavelength pion fluctuations freeze out at the chiral
phase transition. An important aspect revealed by this program is
that  the effective coupling $\lambda_{eff} T \mu^{-\epsilon}$ is
driven to the Wilson-Fischer fixed point in the infrared, this in
turn means that in this limit $\lambda_{eff} \rightarrow 0$. This
may be important in the linear Sigma model description of low
energy QCD near the critical point and may give rise to
interesting phenomenological consequences.

In this article we focused our attention on the approach to the
critical temperature from above, therefore our results regarding
the dynamical exponent $z$ are valid in the symmetric phase. An
important question that we are currently
addressing\cite{boyahecsimio} is the relaxation of pions slightly
below $T_c$. Since the scattering amplitude of pions (at zero
temperature) vanishes in the long-wavelength limit we expect novel
behavior of critical slowing down for pion fluctuations below the
critical temperature.

We expect to report on our findings on these and other related
issues soon\cite{boyahecsimio}.

While we have provided a quantitative implementation of the
program of the $\epsilon$ expansion \emph{with} the resummation
via the renormalization group, the physical limit
$\epsilon\rightarrow 1$ requires higher order calculations with
Borel or Pade resummations much in the same way as in static
critical phenomena. We have studied the dynamical aspects to
lowest order in the $\epsilon$ expansion but clearly a formal
proof of the consistency of the $\epsilon$ expansion to higher
orders, just as in usual critical phenomena, must be explored.

While clearly such programs are beyond the scope and goals of this
article, we here provided the first steps of the program whose
potential phenomenological implications as well as intrinsic
interest in finite temperature quantum field theory warrant
further study.

\vspace{3mm}

\textbf{Acknowledgements.} The work of D.B.\  was supported in
part by the US NSF under grants PHY-9988720 and NSF-INT-9815064,
he thanks the LPTHE at the Universit\'e de Paris VI for
hospitality during this work. H. J. de V. thanks the Department of
Physics at the University of Pittsburgh for hospitality. The
authors thank M. Simionato, S. B. Liao, E. Mottola and D. T. Son
for useful conversations.

\vspace{5mm}

\appendix

\section{\label{sec:appendix1} One-loop diagram at high temperature}

We derive in this appendix the behaviour of the one-loop diagram $
H(p,s) $ contributing to the four points function for high
temperatures $ T \gg p, s $. Setting $v=1$ to avoid cluttering of
notation (the velocity of light is not relevant for the discussion
in this section), we have from eq.(\ref{unlazo})

\begin{equation} \label{unlazoA} H(p,s) =  \; \frac{T}2 \sum_{l \in {\cal Z}}
\int {d^{d}q \over (2\pi)^{d} } {1 \over \left[ q^2 + (2 \pi \; T
\; l)^2 \right]  \left[ ({\vec q}+{\vec p})^2 +(2 \pi \; T)^2 \,
(n+l)^2 \right] } \; , \end{equation} where $ d = 4 - \epsilon $
is the number of spatial dimensions. The denominators in eq.
(\ref{unlazoA}) can be combined using Feynman parameters with the
result \begin{eqnarray}\label{hps2} H(p,s) &=& \frac{T}{2}
\sum_{l \in {\cal Z}} \int {d^{d}q \over (2\pi)^{d} } \int_0^1 dx
\; {1 \over \left[ q^2 + A_l(x,p,s) \right]^2} \cr \cr \cr
 &=& \frac{T}{2(4\pi)^{d\over 2} }
\Gamma\left(2 - {d \over 2}\right) \; \sum_{l \in {\cal Z}}
\int_0^1 dx \; \left[A_l(x,p,s)\right]^{{d \over 2}-2}
\end{eqnarray} where we integrated over the spatial momenta and
 \begin{equation}\label{Aofl}
A_l(x,p,s) = x(1-x) (p^2 + s^2) + (\omega_l + x s)^2 \; ; \;
\omega_l=2\pi T l \, ; \, s=2\pi T n. \end{equation}
  We single out now the contribution from the $ l = 0 $ mode and
study the behaviour of the sum over $l \neq 0 $ for large $ T \gg
p, s $.

Let us first evaluate  the $l = 0 $ term in the sum (\ref{hps2}).
\begin{eqnarray} && \frac{T}{2 (4\pi)^{2-{\epsilon\over 2}}} \;
\Gamma\left({\epsilon \over 2}\right) \;  \int_0^1 dx \;
\left[A_0(x,p,s)\right]^{-{\epsilon \over 2}} =\cr \cr && =
\frac{T}{2 (4\pi)^{2-{\epsilon\over 2}}} \; \Gamma\left({\epsilon
\over 2}\right) \;  \int_0^1 x^{-{\epsilon\over 2}} \; dx \;
\left[ (1-x)p^2 + s^2 \right]^{-{\epsilon \over 2}}=\cr \cr && =
-\frac{T \mu^{-\epsilon}}{2 } {\Gamma\left( {\epsilon
  \over 2}-1\right)
 \over \left( {4 \pi} \right)^{2-{\epsilon \over 2 }}}
 \left(\frac{s^2 + p^2}{ \mu^2} \right)^{ - {\epsilon \over 2}} \;
F\left({\epsilon \over 2},1-{\epsilon \over 2};2-{\epsilon \over
2};{p^2 \over p^2 + s^2}\right) \; , \end{eqnarray} where $
F(a,b;c;z) $ stands for the hypergeometric function\cite{gr}.

We have for the $ l \neq 0 $ terms in the high temperature limit,
\begin{eqnarray} &&\left[A_l(x,p,s)\right]^{-{\epsilon \over 2}}
\buildrel{T \gg p,s}\over = \left( 2 \pi T |l|\right)^{-\epsilon }
- \mbox{sign}(l) \; { \epsilon \, x \, \over |2\pi T
l|^{1+\epsilon }} \cr \cr &&-  { \epsilon \, x \, \over 2 \, |2\pi
T l|^{2+\epsilon }}\left\{ (1-x) p^2 + s^2[1 - (2+\epsilon) x
]\right\} + {\cal O}\left(\left|T \, l \right|^{-3
-\epsilon}\right) \; . \end{eqnarray}

The sum over $ l \neq 0 $ then yields in the high temperature
limit, \begin{eqnarray} &&\frac{ T }{ 2 \,
(4\pi)^{2-{\epsilon\over 2}}} \; \Gamma\left({\epsilon \over
2}\right) \;  \int_0^1 dx \;\sum_{l\neq 0 }
\left[A_l(x,p,s)\right]^{-{\epsilon \over 2}} \buildrel{T \gg
p,s}\over =\cr \cr &&=\frac{\left( 2 \pi T \right)^{1-\epsilon}
\Gamma\left(\frac{\epsilon}{2}\right)}{ 2 \,
(4\pi)^{2-{\epsilon\over 2}}} \;  \; \left\{
\frac{\zeta(\epsilon)}{ \pi} \;
-{\epsilon \; \zeta(2+\epsilon) \over 12 \, \pi \left( 2 \pi
T\right)^{2 } } \left[ p^2 - s^2(1+2 \epsilon) \right] +
 {\cal O}\left(T^{-3 -\epsilon}\right) \right\}
\end{eqnarray} We see that only the first term in the r.h.s. is
important for $ 0<\epsilon< 1 $ and high temperature. This term is
the dominant high temperature limit of the sum of nonzero
Matsubara modes. We then have,
$$
H(p,s)\buildrel{T \gg p,s}\over = H_{asi}(p,s)  -
\frac{\Gamma\left(1 + \frac{\epsilon}2 \right) \;
\zeta(2+\epsilon)}{  384 \; \pi^{4+\frac{\epsilon}2} \; T^{1 +
\epsilon}} \left[ p^2 - s^2(2-\epsilon) \right] \left[ 1 + {\cal
O}\left({p^2 , s^2 \over T^2}\right)\right] \; ,
$$
where \begin{eqnarray} \label{4ptonuA} &&H_{asi}(p,s) \equiv
-\frac{T \mu^{-\epsilon}}{2 } {\Gamma\left( {\epsilon
  \over 2}-1\right)
 \over \left( {4 \pi} \right)^{2-{\epsilon \over 2 }}}
 \left({s^2 + p^2 \over \mu^2} \right)^{ - {\epsilon \over 2}} \;
F\left({\epsilon \over 2},1-{\epsilon \over 2};2-{\epsilon \over
2};{p^2 \over p^2 + s^2}\right) + \cr \cr &&+ {\mu^{2\epsilon-2}
\; \Gamma\left(1 + \frac{\epsilon}2 \right) \; \zeta(\epsilon)
\over 8 \; \pi^{2+\frac{\epsilon}2} \; \epsilon } \; T^{1 -
\epsilon} \; . \end{eqnarray}

\section{\label{sec:appendix2} Two loop diagram at high temperature}

The renormalized two points function is given by (all quantities
are renormalized below)

\begin{equation} \label{beta} \Gamma^{(2)}(p,s) = Z_{\phi} \; p^2 +  {Z_v \over
v^2} \; s^2 -  \Sigma^{(2)}(p,s) +{\cal O}(\lambda^3)
\end{equation}

where $ \Sigma^{(2)}(p,s)$ is the two-loops self-energy.  Using
the renormalization conditions (\ref{renconds}) we find for the
wave function and velocity of light renormalizations,
\begin{eqnarray}\label{zetex} Z_{\phi} &=& 1 +  \; \left. {\partial
\Sigma^{(2)}(p,s) \over
\partial p^2} \right|_{p=\mu, s = \mu \, v}  + {\cal O}(
\lambda^3) \; , \cr \cr Z_v &=& 1 +\left. v^2 \; {\partial
\Sigma^{(2)}(p,s) \over \partial s^2} \right|_{p=\mu, s = \mu\, v}
+ {\cal O}(\lambda^3) \; . \end{eqnarray} The two loops
contribution to the self energy $\Sigma^{(2)}(p,s)$ is given by
\begin{equation}\label{SigmaTA} \Sigma^{(2)}(p,s) = {\lambda^2 T^2
\over 6}\sum_{l,j \in {\cal Z}} \int {d^{4 - \epsilon}q \over
(2\pi)^{4 - \epsilon} } \;{d^{4 - \epsilon}k \over (2\pi)^{4 -
\epsilon} }{1 \over [ q^2 + \frac{\omega^2_l}{v^2}] \; [ k^2 +
\frac{\omega^2_j}{v^2}] \; \left[ ({\vec p}+{\vec k}+{\vec q})^2 +
\frac{(\omega_l+\omega_j+s)^2}{v^2} \right] } \; . \end{equation}
 where $\omega_j=2\pi \, j\, T;  s = 2\pi \, T\, n \, $.

We combine the propagators in eq.(\ref{SigmaTA}) using Feynman
parameters and integrate over the momenta with the result
\begin{eqnarray}\label{sigma2} &&\Sigma^{(2)}(p,s) ={\lambda^2 T^2
~\Gamma(\epsilon-1) \over 6 (4 \pi)^{4-\epsilon}} \sum_{l,j \in
{\cal Z}} \int_0^1 dx \int_0^1 d \xi \; { x^{{\epsilon \over 2}-1}
\over \left[ 1 -x + x \xi(1-\xi) \right]^{2 - {\epsilon \over 2}}}
\\ \cr &&\left\{ { x(1-x) \xi(1-\xi) \over 1 -x + x \xi(1-\xi)}\;
p^2 + \left(\frac{2 \pi T}{v}\right)^2\left[ j^2 \; x\xi + l^2 \;
x(1-\xi) + (j+l+n)^2 (1-x)\right] \right\}^{1-\epsilon} \nonumber
\end{eqnarray} Using the definition (\ref{dimenlessbare}) for  the
dimensionless coupling,  to order $ g^2 $ we find from eqs.
(\ref{zetex}) and (\ref{sigma2}) \begin{eqnarray}\label{zetas}
Z_{\phi} &=& 1 - \frac{g^2}{6} \; \left(\frac{v\mu}{2\pi
T}\right)^{2\epsilon} \; \Gamma(\epsilon) \sum_{l,j \in {\cal Z}}
\int_0^1 dx \int_0^1 d \xi \; \frac{x^{\frac{\epsilon}{2}} (1 -x)
\, \xi(1-\xi)]}{\left[ 1 -x + x \xi(1-\xi) \right]^{3 -
\frac{\epsilon}{2}}} \cr \cr &&\left[ { x(1-x) \xi(1-\xi) \over 1
-x + x \xi(1-\xi)}\; \left({v\mu \over 2 \pi \,T}\right)^2 + j^2
\; x\xi + l^2 \; x(1-\xi) + \left(j+l+{\mu \over 2 \pi \,T}
\right)^2 (1-x) \right]^{-\epsilon}  \; , \cr \cr Z_v  &=& 1 -
\frac{g^2}{3} \; \left({2\pi T \over v\mu}\right)^{1-2\epsilon} \;
\Gamma(\epsilon) \sum_{l,j \in {\cal Z}} \int_0^1 dx \int_0^1 d
\xi \; { x^{{\epsilon \over 2}-1} (1 -x) \, \left(j+l+{v\mu \over
2 \pi \,T}\right) \over \left[ 1 -x + x \xi(1-\xi) \right]^{2 -
{\epsilon \over 2}}} \cr \cr &&\left[ { x(1-x) \xi(1-\xi) \over 1
-x + x \xi(1-\xi)}\; \left({v\mu \over 2 \pi \,T}\right)^2 + j^2
\; x\xi + l^2 \; x(1-\xi) + \left(j+l+{v\mu \over 2 \pi
\,T}\right)^2 (1-x) \right]^{-\epsilon} \; . \end{eqnarray} We
find from the definition of the anomalous dimension
(\ref{gammaexp}) and (\ref{zetas}) for $ \gamma(g,{T \over \mu},v)
$,
\begin{eqnarray}\label{bega2} && \gamma(g,{T \over \mu},v) = g^2 \;
{\Gamma(\epsilon+1) \over 3 } \sum_{l,j \in {\cal Z}} \int_0^1 dx
\int_0^1 d \xi \; { x^{{\epsilon \over 2}}\, (1 -x)^2 \, \xi
(1-\xi) \over \left[ 1 -x + x \xi(1-\xi) \right]^{4 - {\epsilon
\over 2}}} \times  \cr \cr && \left( 1-x + 2 \, x \, \xi(1-\xi)
\left[ 1 + {\pi\, T\over v\mu}(j+l) \right] \right) \times  \\ \cr
&&\left\{ { x(1-x) \xi(1-\xi) \over 1 -x + x \xi(1-\xi)} +
\left({2\pi\, T\over v\mu} \right)^2\left[ j^2 \; x\xi + l^2 \;
x(1-\xi) + \left(j+l+{v \mu \over 2 \pi \,T}\right)^2 (1-x)
\right]\right\}^{-1-\epsilon}  + {\cal O}(g^3) \; . \nonumber
\end{eqnarray} We split the expression for $ \gamma(g,{T \over \mu},v) $
as follows,
$$
\gamma(g,{T \over \mu},v)=\gamma_0(g,v)+\gamma_{nz}(g,{T \over
\mu},v) \; ,
$$
where $ \gamma_0(g,v) $ is the contribution from the zero
Matsubara mode in eq.(\ref{bega2})
$$
\gamma_0(g,v) = g^2 \; {\Gamma(\epsilon+1) \over 3 } \int_0^1 dx
\int_0^1 d \xi \; { x^{{\epsilon \over 2}} \, (1 -x)^{1-\epsilon}
\, \xi \, (1-\xi) \over \left[ 1 -x + x\xi(1-\xi) \right]^{3 -
{3\epsilon \over 2}} \left[ 1 -x + 2\, x\xi(1-\xi)
\right]^{\epsilon}} + {\cal O}(g^3)
$$
and $ \gamma_{nz}(g,{T \over \mu},v) $ stands for the contribution
of the non-zero Matsubara modes.

For $ T\gg \mu $, we see from eq.(\ref{bega2})  that $
\gamma_{nz}(g,{T \over \mu},v) $  decreases as $ \left({T\over
\mu} \right)^{-2-\epsilon} $. [Notice that the coefficient of $
\left({T\over \mu} \right)^{-1-\epsilon} $ vanishes by symmetry
when summing over $j+l$].

Therefore, $ \gamma_0(g,v) $ dominates for $ T\gg \mu . \;
\gamma_0(g,v) $ can be easily computed for small $ \epsilon > 0 $
with the result, \begin{eqnarray}\label{gama4} &&\gamma_0(g,v) =
\frac{g^2}{3} \; \int_0^1 dx \int_0^1 d \xi \; { (1 -x) \, \xi \,
(1-\xi) \over \left[ 1 -x + x\xi(1-\xi) \right]^3} + {\cal
O}(\epsilon \, g^2,g^3) \cr \cr && = {g^2 \over 6  } + {\cal
O}(\epsilon \, g^2,g^3) \end{eqnarray}
 Therefore,
\begin{equation}\label{gamaf} \gamma(g,{T \over \mu},v) =  {g^2 \over 6} +
{\cal O}(\epsilon \, g^2,\; g^3,\; {\mu^2  \over T^2})
\end{equation}

\bigskip

To the lowest non-trivial order in $g$, that is $g^2$ (two-loops),
we find for the function $\beta_v(g,{T \over \mu},v)$
\begin{equation}\label{betava} \beta_v(g,{T \over \mu},v) = - {v \over 2} \;
\gamma(g,{T \over \mu},v) - {v  \over 2} \left( \mu {\partial
\over \partial\mu} - 2 \epsilon\right) \log Z_v  \; .
\end{equation} where the derivatives are now at constant (bare)
$g$.  Using eq.(\ref{zetas}) for $ \log Z_v $ yields
\begin{eqnarray}\label{Wtot} && W \equiv \left( \mu {\partial \over
\partial\mu} - 2 \epsilon\right) \log Z_v = g^2 \;
{ \pi \Gamma(\epsilon) \over 3 } \sum_{l,j \in {\cal Z}} \int_0^1
dx \int_0^1 d \xi \; { x^{{\epsilon \over 2}-1} (1 -x)  \over
\left[ 1 -x + x \xi(1-\xi) \right]^{2 - {\epsilon \over 2}}} \;
Q_{j,l}(x,\xi)^{-\epsilon} \cr \cr &&\left\{\left[ {x \xi(1-\xi)
\over 1 -x + x \xi(1-\xi)} + 1 \right] { (1-x)\epsilon  \over \pi
\;  Q_{j,l}(x,\xi)  } + \left[ 1 + {2\, \epsilon \, (1-x)\over
Q_{j,l}(x,\xi)} \right](j+l) \; {T \over v\mu} \right\}
\end{eqnarray} where,
$$
 Q_{j,l}(x,\xi) \equiv {x  (1-x) \xi(1-\xi) \over 1 -x + x \xi(1-\xi)}
 + \left({2 \pi \,T\over v\mu}\right)^2 \left[ j^2 \; x \, \xi + l^2 \;
 x(1-\xi)\right] + (1-x) \left[ {2 \pi \,T\over v\mu}(j+l) + 1
 \right]^2 \; .
$$
We find in the high temperature limit $ T \gg \mu $ that this
expression is dominated by its zero mode contribution $W_0$
(corresponding to $ j = l = 0 $),
$$
W_0 = g^2 \; { \Gamma(1+\epsilon) \over 3 }  \int_0^1 dx \int_0^1
d \xi \; { x^{{\epsilon \over 2}-1} (1 -x)^{1-\epsilon}  \over
\left[ 1 -x + x \xi(1-\xi) \right]^{2 - {\epsilon \over 2}}}
\left[  {x \xi(1-\xi) \over 1 -x + x \xi(1-\xi)} + 1
\right]^{-\epsilon}
$$
which turns out to be $T$-independent.

The sum of non-zero terms gives a subdominant contribution for $ T
\gg \mu $ and $ \epsilon $ {\bf strictly positive}. We find from
eq.(\ref{Wtot}) after calculation, \begin{eqnarray} &&W_{nz}
\buildrel{T \gg \mu}\over = g^2 \; \left({v\mu \over 2\pi
T}\right)^{2\epsilon} \; {\Gamma(1+\epsilon) \over 3  } \int_0^1
dx \int_0^1 d \xi \; { x^{{\epsilon \over 2}-1} (1 -x)^2 \over
\left[ 1 -x + x \xi(1-\xi) \right]^{2 - {\epsilon \over 2}}}
\times \cr \cr &&\sum_{l,j \in {\cal Z}} {j^2 \over
\left[j^2(1-x+x\xi) +l^2 \, x -2jlx\xi\right]^{1+\epsilon}}
\left[1 + {\cal O}({\mu^2 \over T^2} , \; g)\right] \; .
\end{eqnarray}

For $ 0 < \epsilon \ll 1 $ and for $ T \gg \mu $, $W_0$ and
therefore $W$ are dominated by the pole of $W_0$ at $ \epsilon = 0
$. That is \begin{equation}\label{Wfin} W \buildrel{T \gg \mu, 0 <
\epsilon \ll 1}\over = { 2 g^2 \over 3  \epsilon } + {\cal
O}\left[ \left({\mu \over T}\right)^{2 \epsilon } , \epsilon^0
\right] \; , \end{equation} where we used that $ x^{{\epsilon
\over 2}-1} \buildrel{\epsilon \to 0}\over= {2 \over \epsilon}
\delta(x) $.

Therefore, we find for $ \beta_v(g,{T \over \mu},v) $ from
eqs.(\ref{betava}), (\ref{Wtot}) and (\ref{Wfin}),
$$
\beta_v(g,{T \over \mu},v) \buildrel{T \gg \mu, 0 < \epsilon \ll
1}\over= { v \; g^2 \over 3  }\left[ {1 \over \epsilon } +  {\cal
O}(\epsilon^0 )\right] + {\cal O}\left[\left({\mu \over
T}\right)^{2 \epsilon } \right]\; .
$$

\section{\label{sec:formalproof} Formal proof.}

The formal proof to one loop order begins with the expression
(\ref{unlazoA}) from appendix \ref{sec:appendix1} above.

We now use the identity\cite{lebellac,kapusta}

\begin{equation}\label{identity} I=T\sum_{l =-\infty}^{l=\infty}
\left[A_l(x,p,s)\right]^{-\frac{\epsilon}{2}} = \int_C
\frac{dk_0}{4\pi~i}\left[A(k_0,x,p,s)\right]^{-\frac{\epsilon}{2}}
\coth\left[\frac{k_0}{2T}\right]\end{equation} with
$A(k_0,x,p,s)=A_l(x,p,s;w_l=-ik_0)$ and the contour $C$ is
displayed in figure (\ref{fig:contour}) below. The function
$\left[A(k_0,x,p,s)\right]^{-\frac{\epsilon}{2}}$ has a cut
running parallel to the real axis which for $p \neq 0$ or in the
massive case for arbitrary $p$ begins away from the imaginary axis
and the contour $C$. The contour can now be deformed to wrap
around the cut and the analytic continuation $s\rightarrow
-i\omega +0^+$ can be performed. For $p,\omega \ll T$ the infrared
behavior is dominated by $k_0 \ll T$ for which
$\coth\left[k_0/2T\right] \sim 2T/k_0$ and the resulting
expression features a pole at $k_0 =0$ while the cut begins away
from the origin. The cut can be deformed again to circle the
origin and the integral is simply the residue at the pole $k_0=0$.
Therefore the infrared dominant term is given by $I_{ir}=
\left[A_0(x,p,s=-i\omega+0^+)\right]^{-\frac{\epsilon}{2}}$,
result which coincides with the analysis in terms of the Matsubara
sums provided in appendix \ref{sec:appendix1}.

\begin{figure}[ht!]
\begin{center}
\includegraphics[keepaspectratio=true]{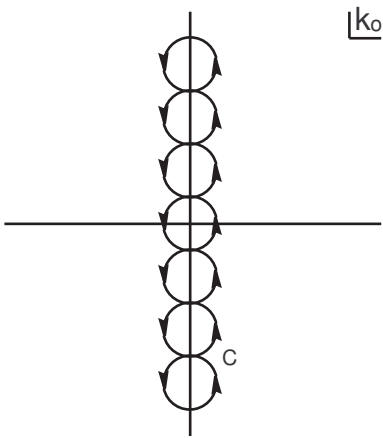}
\caption{Contour in the complex $k_0$ plane} \label{fig:contour}
\end{center}
\end{figure}


\begin{thebibliography}{99}
\bibitem{footnote2} Note that the quartic
self coupling in ref. \cite{relaxationI} has a different
normalization that differs from the one used here by a factor of
12.
\bibitem{footnote3} It is given by $\tilde{\rho}_{cl}$ right before equation (3.8)
in\cite{relaxationI}.
\end{thebibliography}
\end{document}